%% file: GRB190829A_referee.tex
\documentclass[twocolumn]{aastex63}

\usepackage{xspace}

\newcommand{\thisgrb}{GRB~190829A\xspace}
\newcommand{\fermi}{{\em Fermi}\xspace}
\newcommand{\fermiT}{{T$_{0}$}\xspace}
\newcommand{\keV}{{\rm ~keV}\xspace}
\newcommand{\erg}{{\rm ~erg}\xspace}
\newcommand{\s}{{\rm ~s}\xspace}
\newcommand{\swift}{{\em Swift}\xspace}

\newcommand{\tninty}{{T$_{90}$}\xspace}
\newcommand{\Ep}{E$_{p}$}
\newcommand{\sw}[1]{\texttt{#1}}
\graphicspath{{./}{figures/}}
\usepackage{color}
\usepackage{amsmath}

\shorttitle{\thisgrb}
\shortauthors{Vikas Chand et al.}

\begin{document}
\title{Peculiar prompt emission and afterglow in H.E.S.S. detected GRB 190829A}

\author[0000-0002-7876-7362]{Vikas Chand$^\dagger$}
\affiliation{School of Astronomy and Space Science, Nanjing
University, Nanjing 210093, China}
\affiliation{Key Laboratory of Modern Astronomy and Astrophysics (Nanjing University), Ministry of Education, China}
\author[0000-0002-4371-2417]{Ankush Banerjee}
\noaffiliation{}
\author[0000-0003-4905-7801]{Rahul Gupta}
\affiliation{Aryabhatta Research Institute of Observational Sciences (ARIES), Manora Peak, Nainital-263002, India.}
\affiliation{Department of Physics, Deen Dayal Upadhyaya Gorakhpur University, Gorakhpur 273009, India}
\author[0000-0001-9868-9042]{Dimple}
\affiliation{Aryabhatta Research Institute of Observational Sciences (ARIES), Manora Peak, Nainital-263002, India.} 
\affiliation{Department of Physics, Deen Dayal Upadhyaya Gorakhpur University, Gorakhpur 273009, India}
\author[0000-0001-8922-8391]{Partha Sarathi Pal}
\affiliation{School of Physics and Astronomy, Sun Yat Sen University, Guangzhou 510275, China}
\author[0000-0003-3383-1591]{Jagdish C. Joshi}
\affiliation{School of Astronomy and Space Science, Nanjing
University, Nanjing 210093, China}
\affiliation{Key Laboratory of Modern Astronomy and Astrophysics (Nanjing University), Ministry of Education, China}
\author[0000-0003-4111-5958]{Bin-Bin Zhang$^{*}$}
\affiliation{School of Astronomy and Space Science, Nanjing
University, Nanjing 210093, China}
\affiliation{Key Laboratory of Modern Astronomy and Astrophysics (Nanjing University), Ministry of Education, China}
\affiliation{Department of Physics and Astronomy, University of Nevada Las Vegas, NV 89154, USA}
\author[0000-0001-5488-7258]{R. Basak}
\affiliation{INAF-- OAS Bologna, Via P. Gobetti 101, 41129 Bologna, Italy}
\affiliation{Department of Physics and Earth Sciences,
University of Ferrara, Block C, Via Saragat--1, 41122 Ferrara, Italy}
\author[0000-0002-1262-7375]{P. H. T. Tam}
\affiliation{School of Physics and Astronomy, Sun Yat Sen University, Guangzhou 510275, China}
\author[0000-0002-4394-4138]{Vidushi Sharma}
\affiliation{Inter University Centre for Astronomy and Astrophysics, Pune, India}
\author{S. B. Pandey}
\affiliation{Aryabhatta Research Institute of Observational Sciences (ARIES), Manora Peak, Nainital-263002, India.}
\author{Amit Kumar}
\affiliation{Aryabhatta Research Institute of Observational Sciences (ARIES), Manora Peak, Nainital-263002, India.}
\affiliation{School of Studies in Physics and Astrophysics, Pandit Ravishankar Shukla University, Chattisgarh 492 010, India}
\author[0000-0002-7555-0790]{Yi-Si Yang}
\affiliation{School of Astronomy and Space Science, Nanjing
University, Nanjing 210093, China}
\affiliation{Key Laboratory of Modern Astronomy and Astrophysics (Nanjing University), Ministry of Education, China}
\email{$^\dagger$vikasK2@nju.edu.cn}
\email{$^{*}$bbzhang@nju.edu.cn}

\begin{abstract}
We present the results of a detailed investigation of the prompt and afterglow emission in the HESS detected  \thisgrb.
\swift and \fermi observations of the prompt phase of this GRB reveal two isolated sub-bursts or episodes, separated by a quiescent phase. The energetic and the spectral properties of the first episode are in stark contrast to the second. The first episode, which has a higher spectral peak $\sim 120 \keV$ and a low isotropic energy $\sim 10^{50}\erg$ is an outlier to the Amati correlation and marginally satisfies the Yonetoku correlation. However, the energetically dominant second episode has lower peak energy and is consistent with the above correlations. We compared this GRB to other low luminosity GRBs (LLGRBs). Prompt emission of LLGRBs also indicates a relativistic shock breakout origin of the radiation. For \thisgrb, some of the properties of a shock breakout origin are satisfied. However, the absence of an accompanying thermal component and energy above the shock breakout critical limit precludes a shock breakout origin.
In the afterglow, an unusual long-lasting late time flare of duration $\sim 10^4 \s$ is observed. 
We also analyzed the late-time \fermi-LAT emission that encapsulates the HESS detection. 
Some of the LAT photons are likely to be associated with the source.
All above observational facts suggest \thisgrb is a peculiar low luminosity GRB that is not powered by a shock breakout, and with an unusual rebrightening due to a patchy emission or a refreshed shock during the afterglow.  Furthermore, our results show that TeV energy photons seem common in both high luminosity GRBs and LLGRBs.

\end{abstract}

\section{Introduction} \label{sec:intro}
The radiation mechanisms in the prompt emission of the gamma-ray bursts (GRBs) remain a highly debated topic. On the other hand, the afterglow phase, in general, is well explained by the emission originating from external shocks produced by a blastwave inevitably crashing into the circumburst medium, and any deviations from this model can also be addressed (e.g., see \citealt{Kumar:2015PhR, Meszaros:2019MmSAI} for a review). 

The recent detections of GRB afterglow in TeV energies \footnote{several hundred GeVs} by H.E.S.S. and MAGIC Cerenkov telescopes have provided new insights in the  
study of GRBs \citep{Abdalla:2019Natur, MAGIC2019Natur_detection, HESS:gcn25566}. For example, GRB 190114C, in its multifrequency spectral energy density (SED) showed evidence for a double-peaked distribution with the first peak being the synchrotron emission. The second peak shows a very high energy (VHE) emission in TeV energies and is explained by the synchrotron self-Comptonisation process, theoretically predicted in a standard afterglow model. \citep{MAGIC2019Natur_IC}. GRB 180720B also showed a VHE emission at late times that could be explained by the inverse Compton mechanism \citep{Abdalla:2019Natur}.

While the afterglow studies have progressed considerably, the prompt emission is still challenging to understand. A lot of empirical models have been proposed, which include traditional Band function \citep{Band:1993} and deviations from this simple shape modeled by adding an extra thermal component, breaks or multiple spectral components \citep{Ryde:2005, Abdo:2009, Page:2011, Guiriec:2011, Ackermann:2013, Guiriec:2015ApJ, Guiriec:2015b, BR:2015, Guiriec:2016ApJ, Vianello:2017, Ravasio:2018A}.
The recent developments in the physical modeling of the prompt emission show that the synchrotron could be the main emission mechanism \citep{Oganesyan:2019, Burgess:2019NatAs}. However, the physical photospheric models also equally well explains the data \citep{Vianello:2017, Ahlgren:2019ApJ, Zeynep:2020arXiv}. In this context, it is important to study the prompt emission properties of the GRBs detected by the HESS and MAGIC telescopes to get a global picture and capture the diversity of these events. 

Studies on 
GRB 190114C showed multiple components in its prompt emission 
and a standard afterglow \citep{Wang:2019arXiv, Chand:2019arXiv, Fraija:2019ApJ, Ravasio:2019}.
GRB 180720B has synchrotron spectrum for prompt emission and a standard afterglow \citep{Ronchi:2019arXiv, Fraija:2019arXiv1905}. 
\thisgrb is another such GRB detected by the HESS at a redshift of 0.0785 \citep{HESS:gcn25566, GTC:gcn25565}. Compared to the previously detected VHE events, it has a lower luminosity. Prompt emission of LLGRBs also indicates a relativistic shock breakout origin of the radiation \citep{Nakar:2012ApJ}. Here we report the spectral and temporal analysis of the \emph{Neil Gehrels Swift Observatory} and \emph{Fermi Gamma-ray Space Telescope} data and multiwavelength observations of this event. 

\section{Prompt Observations and Analysis} \label{sec:prompt_emission}
\thisgrb triggered  \fermi/Gamma-ray Burst Monitor (GBM) at 2019-08-29 19:55:53.13 UTC \citep[\fermiT][]{Fermi_GBM:gcn25575} and \swift/Burst Alert Telescope (BAT) at 19:56:44.60 UTC \citep{Swift_BAT:gcn25579}. The \swift/X-ray telescope (XRT) observed 
the GRB from 97.3 $\s$ after the BAT trigger time 
and refined the location 
to RA (J2000): 02$\rm h$ 58$\rm m$ 10.57$\rm s$ and DEC (J2000): -08$\rm d$ 57$\rm '$ 28.6$\rm "$ 
\citep{Swift:gcn25552}. H.E.S.S. 
detected TeV signal 4.2 $\rm hrs$ after the prompt emission in a direction consistent with this location. In a multi-wavelength observation campaign,
\thisgrb was followed by 
several optical, NIR and radio telescopes\footnote{\url{https://gcn.gsfc.nasa.gov/other/190829A.gcn3}} (Section \ref{sec:multiwavelength}). 

During the prompt emission, both \fermi and \swift detected two episodes, the first episode starting from \fermiT to \fermiT + 4 $\s$ followed by a brighter episode from \fermiT + 47.1 $\s$ to \fermiT + 61.4 $\s$. The spectrum of the first episode in the \fermi data is best described by a powerlaw with an exponential high-energy cutoff function having an index of -1.41 $\pm$ 0.08, and a cutoff energy corresponding to a peak energy, \Ep = 130 $\pm$ 20 $\keV$ \citep{Fermi_GBM:gcn25575}.  Whereas the second episode is best fit by a Band function \citep{Band:1993} with \Ep = 11 $\pm$ 1 $\keV$, $\rm \alpha$ = -0.92 $\pm$ 0.62 and $\rm \beta$ = -2.51 $\pm$ 0.01. The observed fluence is 1.27 $\pm$ 0.02 $\times$ 10$^{-5}$ $\rm erg$ $\rm cm^{-2}$ in the 10 - 1000 $\keV$ band with the episodes combined \citep{Fermi_GBM:gcn25575}. From the preliminary spectral results, reported in GCNs, we note the different nature of the two episodes.

In our analysis of \fermi-GBM data, we identified NaI detector numbers 6 and 7 (n6 and n7) by visually examining count rates and with observing angles $<$ $\rm 50^\circ$ to the source position. The angle constraints are to avoid the systematics arising due to uncertainty in the response at larger angles. Among the BGO detectors, BGO 1 (b1) is selected as it is closer to the direction of the GRB. The time-tagged event (TTE) data was reduced using \emph{Fermi Science Tools} software \sw{gtburst}\footnote{\url{https://fermi.gsfc.nasa.gov/ssc/data/analysis/scitools/gtburst.html}}.
We used \sw{XSPEC} \citep{Arnaud:1996} to model the spectrum.
The Bayesian information criteria (BIC) is calculated for each model from the \sw{pgstat} value \citep{Kass:1995}. In all these models used, the power-law model has the least BIC. The \swift-BAT spectrum is obtained
in the BAT mission elapsed times (METs) corresponding to the times of our selection for joint
spectral analysis. The recipe followed for reducing the spectrum is as described in
\swift-BAT software guide\footnote{\url{: http://swift.gsfc.nasa.gov/analysis/
bat_swguide_v6_3.pdf}}. We use \sw{HEASOFT} software version-6.25 with latest calibration database\footnote{\url{https://heasarc.gsfc.nasa.gov/FTP/caldb/}}. We applied gain correction
using bateconvert, then batbinevt was utilised to produce spectrum after making a detector plane image
(dpi), retrieving problematic detectors, removing hot
pixels and subtracting the background using \sw{batbinevt}, \sw{batdetmask}, \sw{bathotpix} and \sw{batmaskwtevt}, respectively. Additionally,
\sw{FTOOLS} \sw{batupdatephakw} and \sw{batphasyserr}
are used for compensating the observed residual in the
responses and for making sure that we have the position of the burst in instrument coordinates. We have
generated the detector response matrix (DRM) using
\sw{batdrmgen}. For a joint analysis of BAT-GBM data (See Figure \ref{fig:Amati_Yonetoku} \& \ref{fig:Amati_Yonetoku_full} and Table \ref{tab:sub_bursts_properties}), the GBM data are grouped to result in minimum 20 counts and $\chi^2$-statistics is optimized for finding the best fit parameters. All quoted errors on spectral parameters correspond to $1 \sigma$ (nominal 68\%).

\subsection{The peculiar nature of the episodes}\label{sec:correlations}
\subsubsection{Light curve and spectrum}

\label{subsec:time_int}
\begin{figure}[htb!]
\includegraphics[scale=0.25]{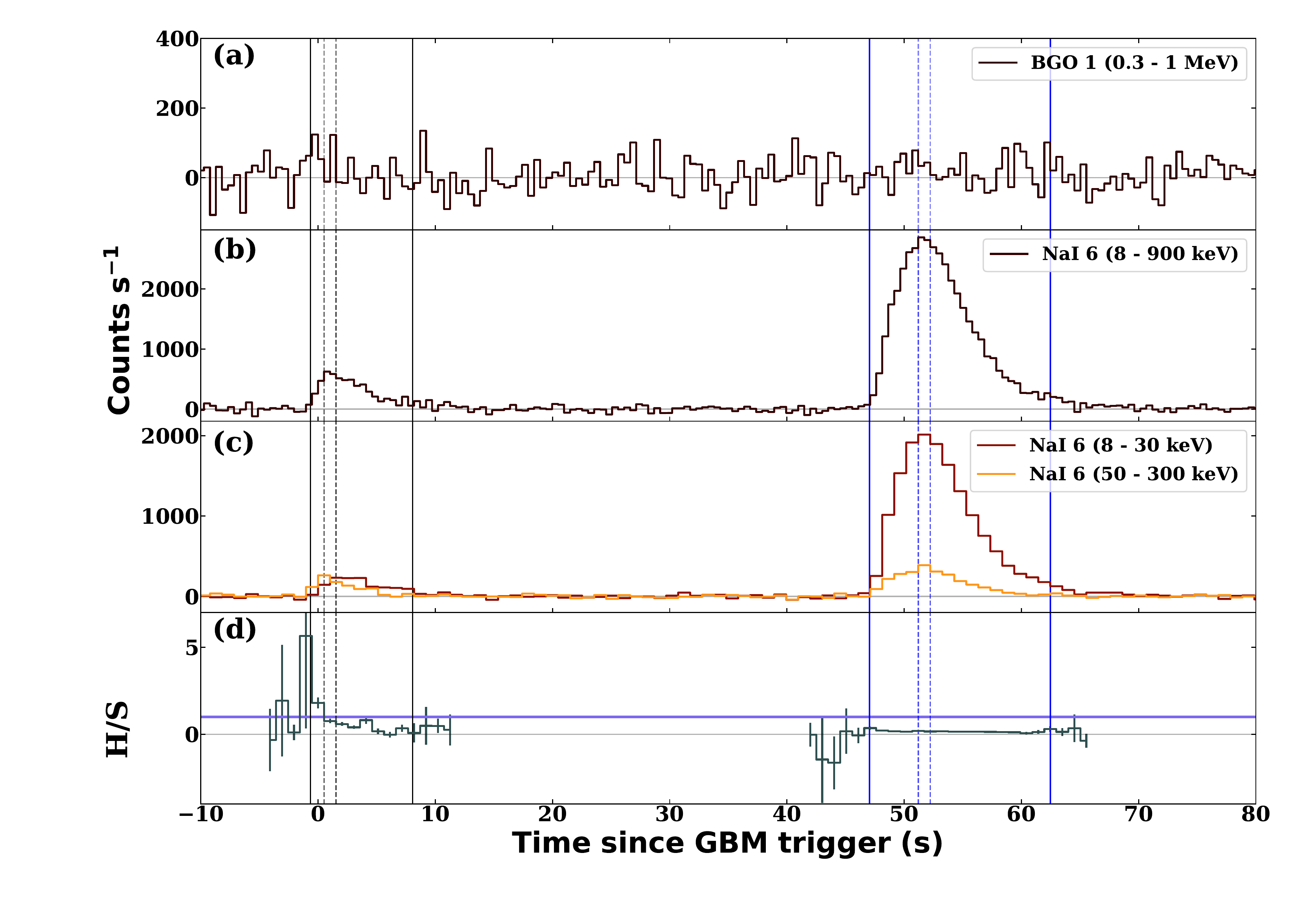}
\caption{{\bf Multi channel Light curves :} The background-subtracted count rate of \thisgrb  with time in multiple energy bands (a -- c).
{\bf Hardness ratio:} (d) The hardness ratio (H/S)  
in energy bands 50 - 300 $\keV$ (H)  and   8 - 30 $\keV$ (S) is shown. The horizontal violet line indicates equal rates in H and S bands. The vertical black and blue lines indicate the boundaries of \tninty (solid lines) and peak luminosity (dashed lines) calculations for the first and second emission episode, respectively.}
\label{fig:HR}
\end{figure}

In Figure \ref{fig:HR}, we have shown the light curves of the prompt emission phase in a wide energy band -- 8 - 900 $\keV$ of GBM-NaI, 0.3 - 1 $\rm MeV$ of GBM-BGO.
The GRB appears to have a softer spectrum, as can be inferred from the low signal of the BGO light curve.
To apprehend it further, we plotted hardness ratio (H/S) in two bands of NaI light curve, where the harder band is 50 - 300 $\keV$, and the softer band is 8 - 50 $\keV$.
We note that the first episode has comparable count rates in these two energy bands, while the second episode has a much higher rate in the softer band, implying a relatively softer nature of this episode.
This is also reflected in the time-integrated spectrum of the individual episodes. Spectral analysis shows that the first episode can be modeled by a power-law (index $\rm \alpha$) with an exponential cutoff, 
where the cutoff energy ($\rm E_{c}$)
can be re-parameterized in terms of peak energy $\rm E_{p}$ = (2+$\alpha$) $\rm E_c$.
The second episode, when modeled with a simple power-law, has a steeper spectral index. 
The properties calculated from spectral parameters for the two episodes are reported in Table \ref{tab:sub_bursts_properties}.

\begin{table}
\begin{scriptsize}
\caption{Properties of two episodes in \thisgrb. Spectral lags are measured at 8 ms bin and  w.r.t 150 - 300 $\rm keV$.}
\label{tab:sub_bursts_properties}
\begin{center}
\begin{tabular}{|c|c|c|}
\hline
Properties & Episode 1 & Episode 2  \\
\hline
\tninty (s) in 50 - 300 $\keV$ & $6.30\pm 0.08$& $10.37 \pm 0.05$ \\ \hline
$\rm t_q$ ($\s$) & \multicolumn{2}{c|}{ $\rm \sim40$}  \\
\hline
HR  & 0.57 & 0.15 \\
$\rm E_{p}$ ($\keV$) &$120.0_{-37.2}^{+111.7}$ & $10.9_{-0.6}^{+0.5}$  \\
$\rm F$ ($\rm 10^{-6} erg ~cm^{-2}$) & $2.4_{-0.4}^{+0.7}$& $13.8_{-0.4}^{+0.4}$ \\
$\rm E_{\gamma, iso}$ ($\rm erg$) & $3.2 \times 10^{49}$ & $1.9\times 10^{50}$\\
$\rm L_{p, iso}$ ($\rm erg ~s^{-1}$) & $8.6\times10^{48}$ & $2.9\times10^{49}$ \\

\hline
Redshift $\rm z$ & \multicolumn{1}{c}{0.0785$\pm$0.005} &\\ \hline

Energy band & \multicolumn{2}{c|}{lag (Correlation)}  \\
($\keV$)& \multicolumn{2}{c|}{$\rm ms$ (\%)}  \\
\hline
8 - 30 & $71\pm3~(63)$ & $100\pm4~(65)$ \\
8 - 100 & $81\pm2~(66)$ & $40\pm3~(66)$ \\
\hline
\end{tabular}
\end{center}
$\rm T_{90}$: Duration from GBM data; $\rm t_q$: quiescent time; HR: ratio of the counts in 50 - 300 $\keV$ to the counts in 10 - 50 $\keV$; $\rm E_{p}$: Time-integrated peak energy calculated using joint BAT (15 - 150 keV) and GBM (8 keV - 40 MeV) data;
$\rm F$: Energy fluence;
$\rm E_{\gamma, iso}$: Isotropic energy; $\rm L_{p, iso}$: Isotropic peak luminosity \\
\end{scriptsize}
\end{table}


We resolved the 8 - 900 $\rm keV$ light curve into smaller bins based on signal to noise ratio (SNR) to study the spectral evolution. 
We created a total of 21 spectra: 5 (SNR = 15) and 16 (SNR = 30) for the first and the second episode, respectively. The cutoff in the power-law model (COMP) is also preferred in the time-resolved analysis of the first episode based on the BIC values, see Table 
\ref{tab:tr_spectral}.
We also see that the spectrum softens with time. The spectral index is $< -2/3$ (within the synchrotron slow cooling limit), and therefore the emission of the first episode can have synchrotron origin. The spectra of the second episode are best fitted by a power-law function with indices $<$ -2. The index is plausible to be related to the higher energy power-law of the Band function. 
When modeled with the Band function, the peak energies are found to be near the lower edge of the \fermi spectral window (i.e. $\rm \sim 8 ~keV$), and hence are probably unphysical as only a few channels are available for the determination of $\alpha$, e.g.,  in case of GRB 171010A \citep{Ravasio:2019AA}.
This is also reflected in erratically changing $\alpha$, which remained unconstrained throughout, see Table \ref{tab:tr_spectral}.

\subsubsection{Amati and Yonetoku correlations}
The spectral peak energy of GRBs in the cosmological rest frame ($E_{\rm p,z}$) is correlated to the isotropic equivalent energy ($\rm E_{\gamma,\rm iso}$) and isotropic peak luminosity ($\rm L_{\gamma,\rm iso}$) in the $\gamma$-ray band, see \cite{Amati:2006MNRAS, Yonetoku:2004ApJ}. 
Amati correlation is also valid for pulse-wise sample of GRBs  (\citealt{Basak:2013MNRAS})
GRB 980425B, GRB 031203A, and GRB 171205A do not satisfy the Amati correlation. GRB 061021 is a notable outlier to both the correlations \citep{Nava:2012MNRAS}.
These correlations have been used to classify individual episodes in GRBs with long quiescent phases \citep{Zhang:2018NatAs}.
We consider these correlations for \thisgrb, in its two episodes of activity.
To check whether the Amati correlation is also followed by individual episodes of two-episode GRBs with a quiescent phase,
we chose the sample of 101 GRBs from \cite{Lan:2018ApJ}.
Among these, there are 11 GRBs with known redshift 
are plotted in Figure~\ref{fig:Amati_Yonetoku} (a \& b). For the rest, the redshift is varied from 0.1 to 10 and their tracks in the correlation plane are studied. Interestingly, all the individual episodes fall within $\rm 3 \sigma$ intrinsic dispersion of the corresponding correlations (see Appendix \ref{sec:two_phases} for the tracks).
But, the first hard-episode of \thisgrb is an outlier to the Amati correlation and marginally satisfies the Yonetoku correlation.

\subsubsection{Hardness ratio (HR) vs \tninty, Spectral lag}
\begin{figure*}[ht]
\centering
 \includegraphics[scale=0.4]{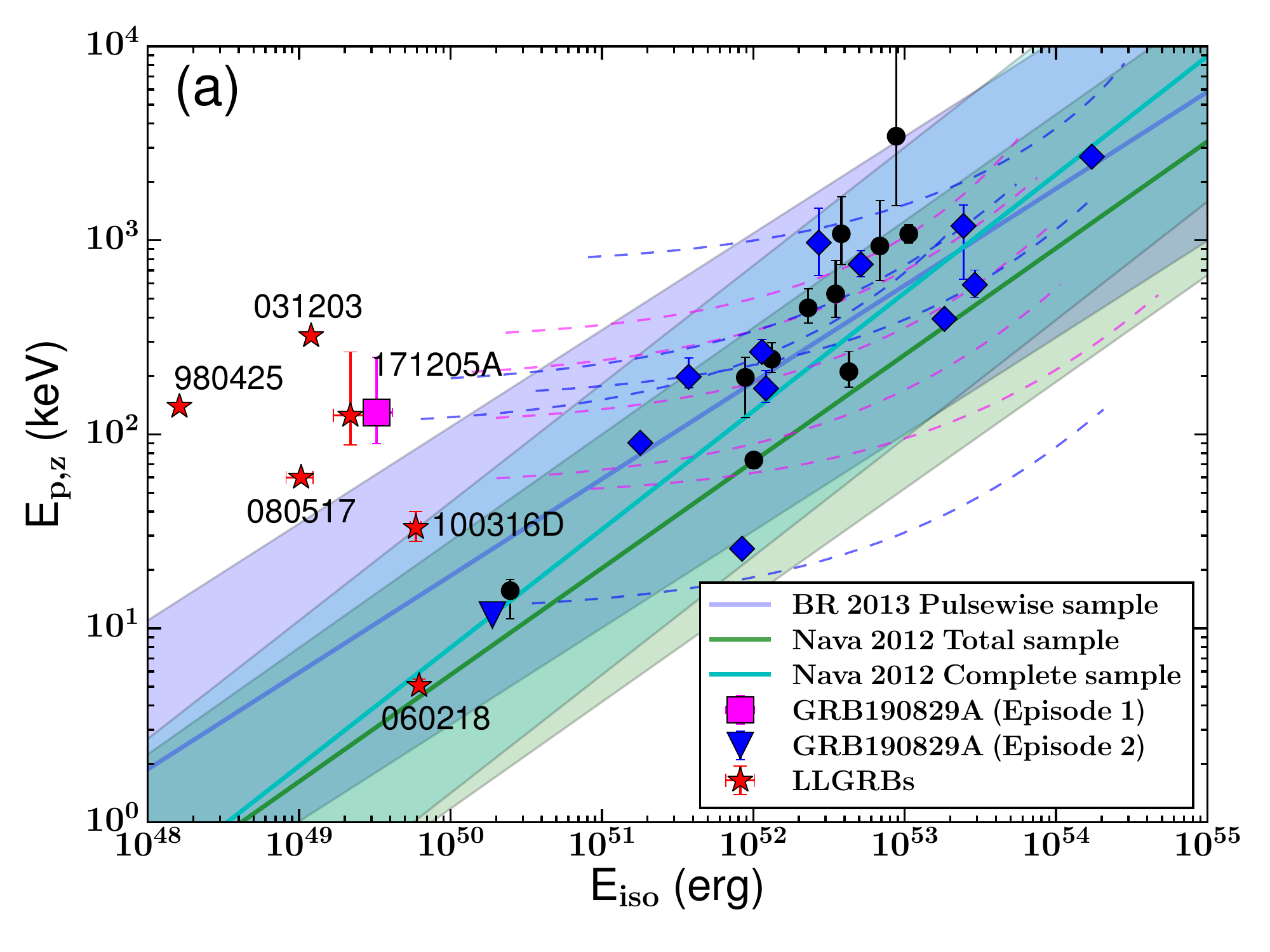}
 \includegraphics[scale=0.4]{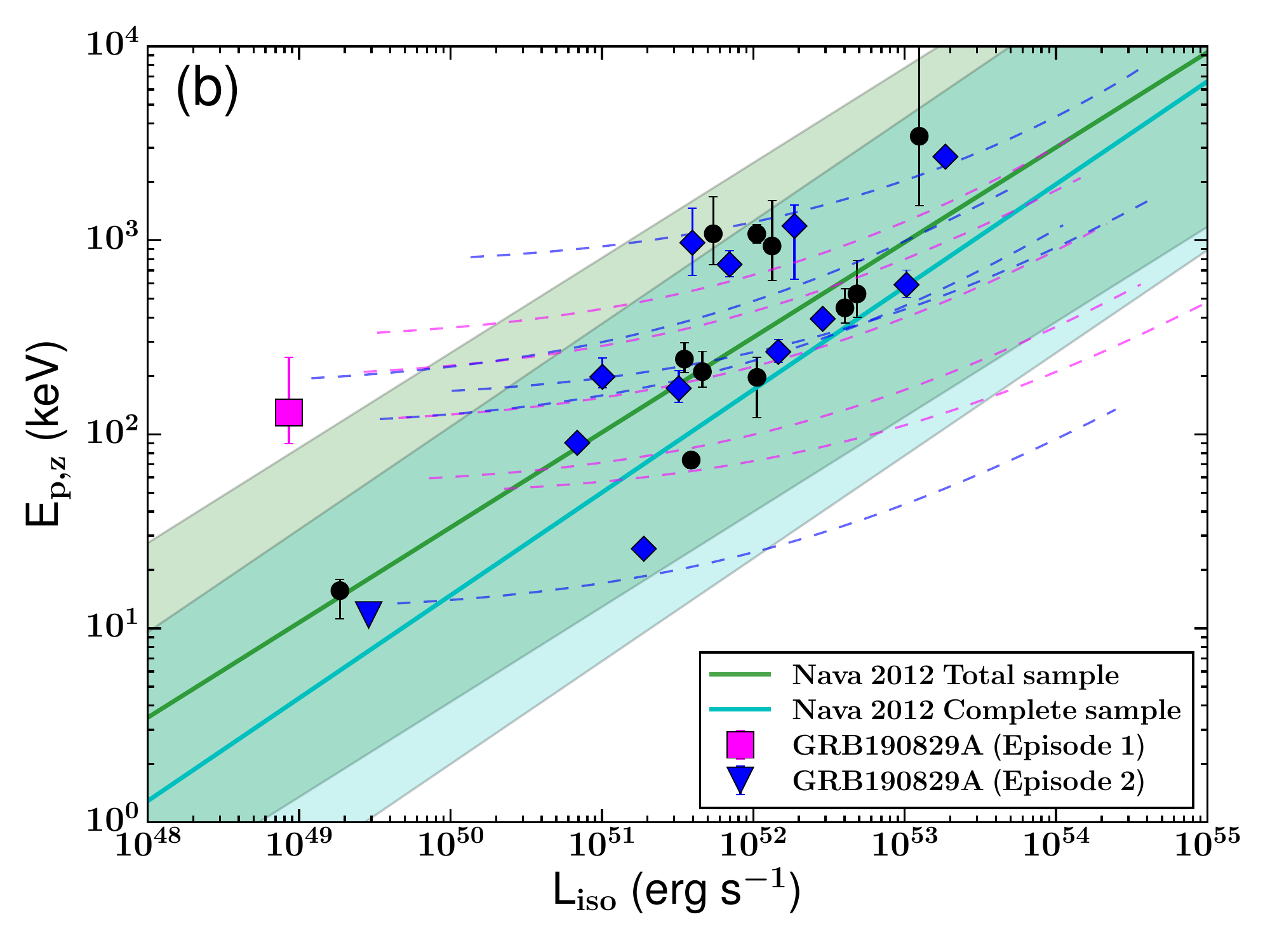}
  \includegraphics[scale=0.4]{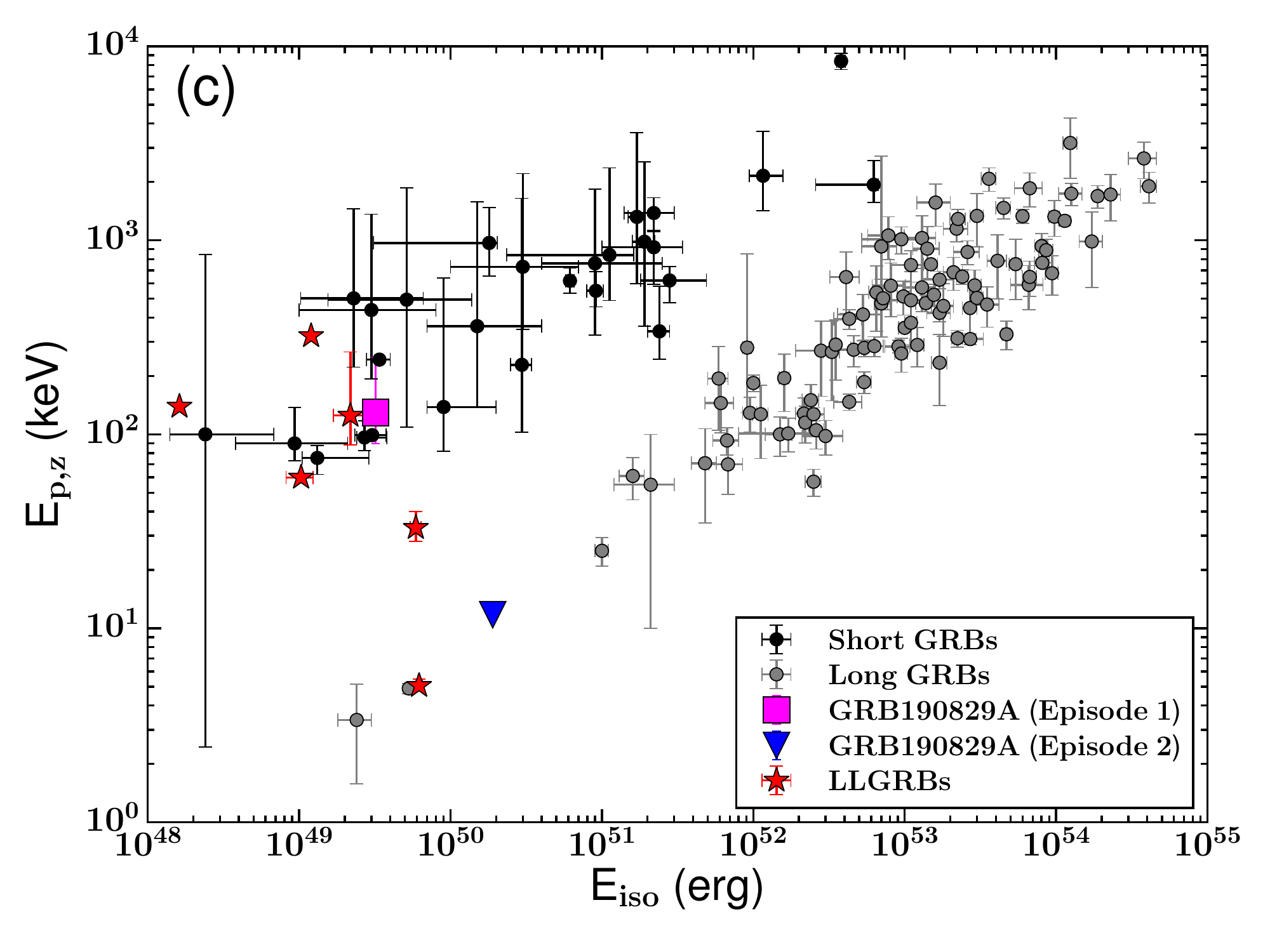}
  \includegraphics[scale=0.4]{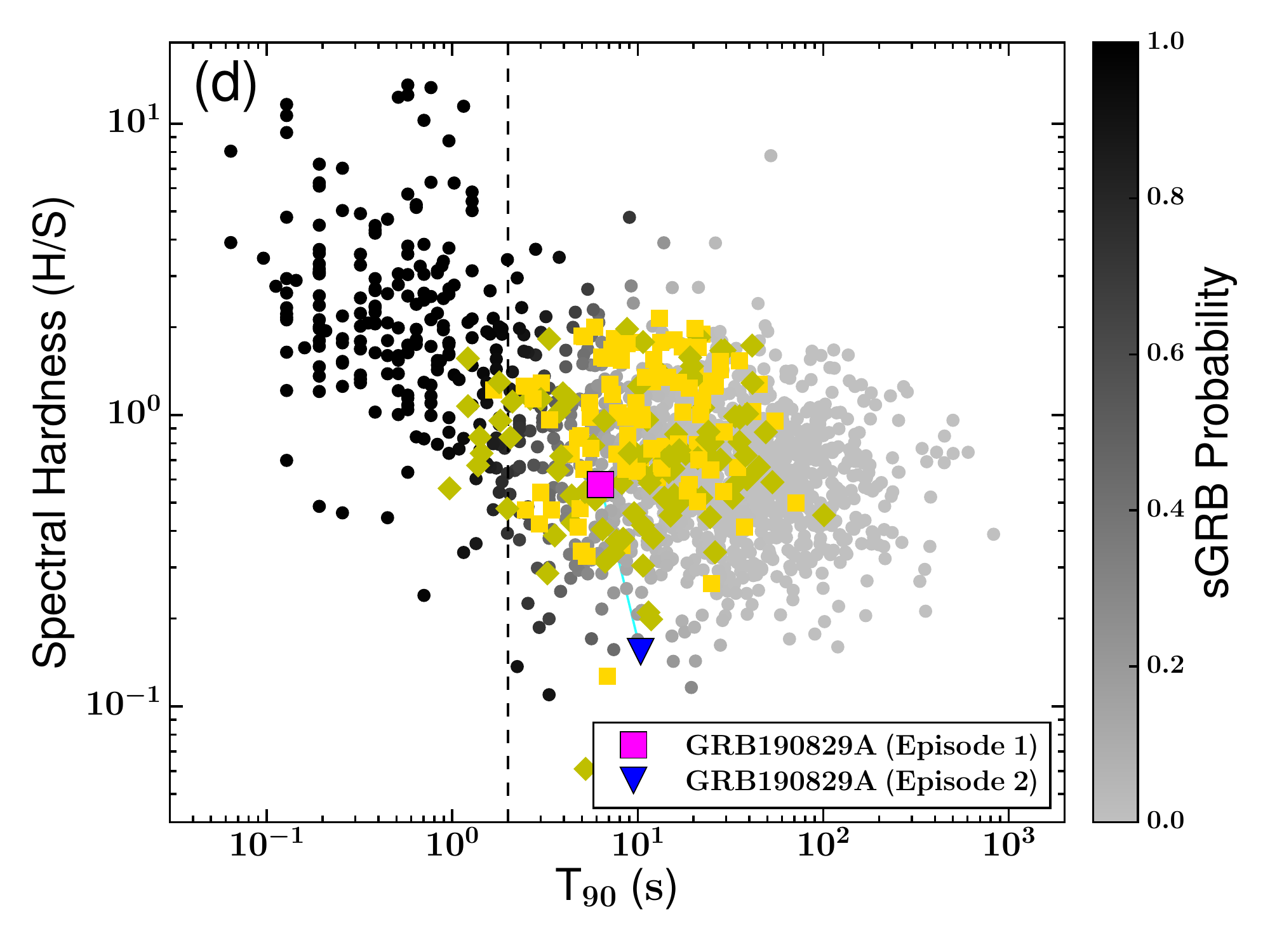}
\caption{{\bf Correlations for two-episode GRBs:} The first episodes (black circles) and second episodes (blue diamonds) of two-episode long GRBs with known redshifts in (a) Amati and (b) Yonetoku correlation plane. GRBs with unknown redshifts are represented by tracks obtained by varying the redshift (For all the tracks, see Appendix \ref{sec:two_phases}). The shaded region represents the 
3$\sigma$ scatter of the correlations  \citep{Nava:2012MNRAS, Basak:2013MNRAS}. The two episodes of \thisgrb are shown in colored symbols. (c) The two episodes of \thisgrb in the Amati correlation plane of long and short  GRBs.  {\bf Hardness ratio:} (d) The spectral hardness and duration \tninty for the two-episode GRBs shown along with the data points for short (black circles) and long GRBs (grey circles) used in \cite{Goldstein:2017ApJ}. Gold squares and yellow diamonds represent the first and second episodes of two-episode long GRBs with known redshifts, respectively.  The color scale represents the probability of a GRB being short (black) or long (grey).}
\label{fig:Amati_Yonetoku}
\end{figure*}

Short GRBs do not follow the same trend in the Amati correlation as the long GRBs. Here we investigate the intriguing possibility that the two episodes of \thisgrb show the properties of the two classes of GRBs.   In Figure~\ref{fig:Amati_Yonetoku} (c), we show the position of the two episodes in the Amati correlation plane of short and long GRB population \citep{Zhang:2018Nat}. Interestingly,  the first episode lies with the short GRB population. 
Classification of long and short GRBs is conventionally studied using their distribution in the hardness-duration plane. The duration, 
\tninty, is calculated by the time period when 5 \% to 95 \% of the total photon fluence is accumulated. We obtained the episode-wise time-integrated HR by dividing the counts in 10 - 50 $\keV$ and $\rm 50$ - $\rm 300\,keV$ energy bands 
to make a comparison with other \fermi GRBs also used in \cite{Goldstein:2017ApJ}. The errors in \tninty and HR are calculated by simulating 10,000 lightcurves by adding a Poissonian noise with the mean values at observed errors \citep{2014AstL...40..235M,Bhat:2016ApJS}.
The \tninty and HR values for \thisgrb are presented in Table \ref{tab:sub_bursts_properties}.
In  Figure \ref{fig:Amati_Yonetoku}(d), we show the HR-\tninty diagram of the two-episode GRBs, with each episode considered separately. The probabilities of a GRB classified as a short or long GRB from the Gaussian mixture model in the logarithmic scale are also shown in the background (taken from \citealt{Goldstein:2017ApJ}). We note that all these data are clustered towards the long GRB category. 
The probability of the first episode being associated with long GRB properties is $\sim87 \%$.

Long GRBs show soft lag where the light curve in low energy band lags behind the lightcurve in high energy band \citep{Fenimore:1995ApJ}, however, many short GRBs do not show statistically significant lag \citep{Bernardini:2015MNRAS}. We calculate the spectral lags for \thisgrb using the discrete cross-correlation function (DCCF) as defined in \cite{Band:1997ApJ}. 
The peak of the observed CCF versus spectral lag is found by fitting an asymmetric Gaussian function \citep{Bernardini:2015MNRAS}.
The lags are calculated between 150 - 300 $\keV$ and the lower energy bands (8 - 30 keV and 8 - 100 keV), and the values are reported in Table \ref{tab:sub_bursts_properties}. The upper value of energy is restricted to 300 $\keV$ because the signal above this energy is consistent with the background.  We chose lightcurves of different resolutions (4, 8, and 16 ms), and the maximum correlation is obtained for  $8$ $\rm ms$. The lags with energy bands and the maximum value of correlations are reported in Table \ref{tab:sub_bursts_properties}. A positive lag is obtained for both the episodes, and it is consistent with the soft lags generally seen in long GRBs.

This analysis through the properties in the hardness ratio vs. T90 diagram and a positive spectral lag for both the episodes strongly suggests that \thisgrb is consistent with the population of long GRBs through a contrasting nature of its episodes in the prompt emission energy correlations.


\section{Multiwavelength modelling}
\label{sec:multiwavelength}

\subsection{H.E.S.S. and \emph{Fermi}-LAT observations}

We extracted the LAT data 
within a temporal window extending 50,000 $\rm s$ after \fermiT. We performed an unbinned likelihood analysis. The data were filtered by selecting photons with energies in the range 100 $\rm MeV$  - 300 $\rm GeV$, within a region of interest (ROI) of $\rm 12^{\circ}$ centered on the burst position. A further selection of zenith angle ($\rm  100^{\circ} $) was applied in order to reduce the contamination of photons coming from the Earth limb. We adopted the P8R3\_SOURCE\_V2 response, which is suitable for longer durations  ($\rm \sim 10^{3} ~s$). The probability of the photons to be associated with \thisgrb is calculated using \sw{gtsrcprob} tool.

\begin{figure}
    \centering
    \includegraphics[scale=0.34]{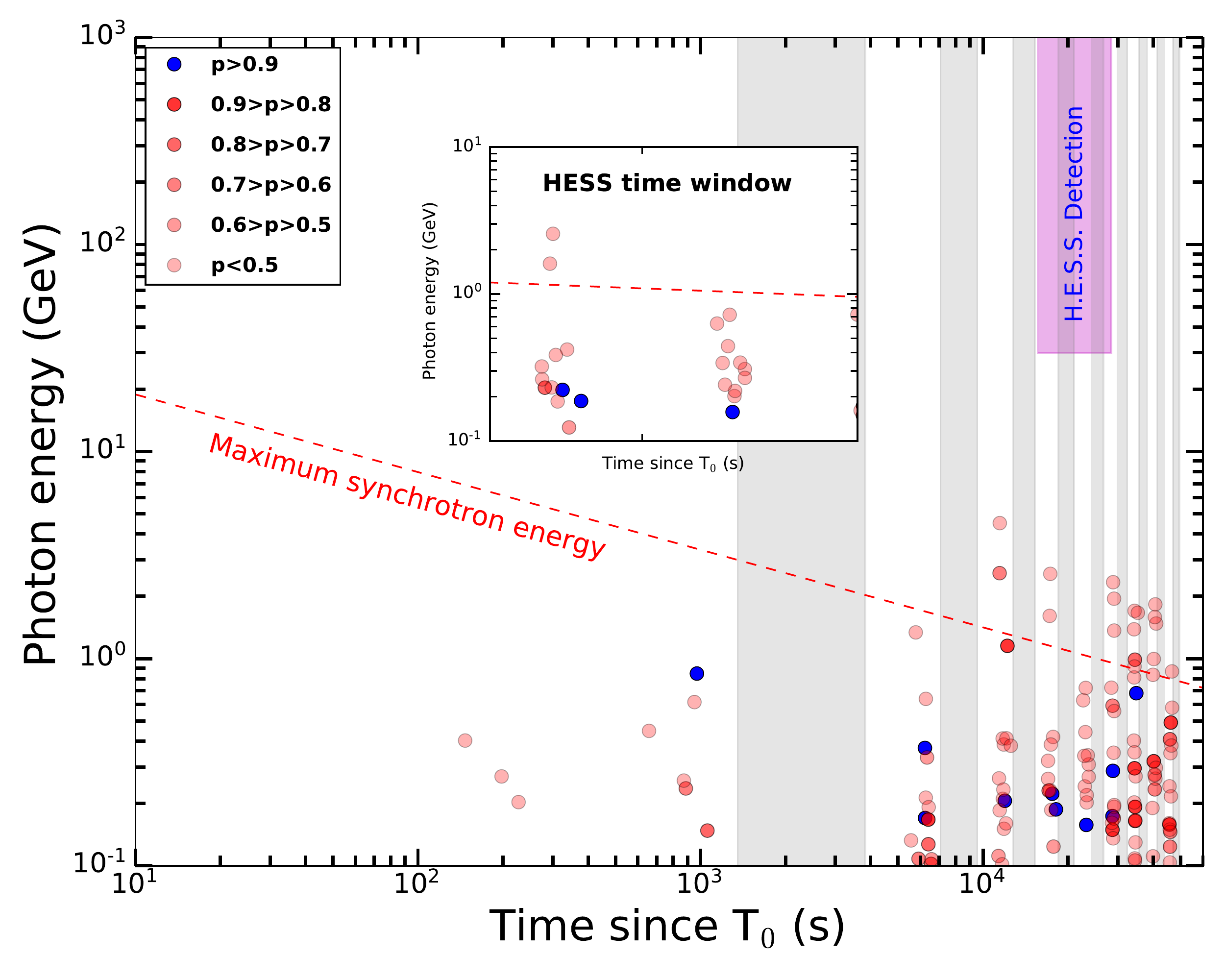}
    \caption{{\bf Delayed emission in LAT:} All the photons with energies $> 100 $ MeV and  probabilities  of being associated with GRB 190829A. The regions with zenith angle $ > 100^\circ$ are shaded grey. 
    The red line represents the maximum photon energies allowed for a synchrotron forward-shock model with an emission efficiency $\rm \eta$ = 1.3\%. The inset shows the LAT emission during the H.E.S.S. time window.}
\label{fig:LAT_burst_fermi}
\end{figure}

We analysed \fermi-LAT data up to  $\rm 5 \times 10^{4} ~s$ after the GBM trigger time, see Figure \ref{fig:LAT_burst_fermi}. We obtained an upper limit on photon flux of $\rm 2.81\times 10^{-7} $   $\rm photons~ cm^{ -2 }s^{-1}$ in $100~\rm MeV$ - $300 ~\rm GeV$. \fermi-LAT detected no photons during the GBM observation, which is consistent with the extrapolated Comptonized spectrum peaking at $\sim$ 114 $\rm keV$ (See Table \ref{tab:sub_bursts_properties}). During the H.E.S.S. observation which started 4.2 $\rm hrs$ after the prompt emission, only three photons are observed in LAT above 100 $\rm MeV$  with probability $ > 90\% $ of their association with the source, though more photons are observed with $>50\%$ probability.

To investigate the origin of the LAT photons, we calculated the maximum photon energy radiated
by the synchrotron process during the deceleration phase \citep{Piran:2010ApJ, Duran:2011MNRAS, Fraija:2019ApJ}. 
The red-dashed line represents the maximum photon energies released by the synchrotron forward-shock model with an emission efficiency of prompt emission $\sim\eta$ = 1.3\% (Section \ref{sec:XRT_analysis}). The LAT photons lying below this line are consistent with synchrotron emission. However, the H.E.S.S. detection would lie above this line
and might be originated due to the synchrotron self-Compton mechanism similar to GRB 190114C and GRB 180720B \citep{MAGIC2019Natur_IC, Abdalla:2019Natur}.

\subsection{X-rays and Optical data}
The \swift X-ray telescope \citep[XRT;][]{2005SSRv..120..165B} began observing the BAT localization field to search for an X-ray counterpart of \thisgrb at 19:58:21.9 UT, 97.3 $\rm s$ after the BAT trigger. The XRT detected a bright and uncatalogued X-ray afterglow candidate at RA (J2000) and DEC (J2000) of 02$\rm h$ 58$\rm m$ 10.57$\rm s$ and -08$\rm d$ 57$\rm '$ 28.6$\rm "$, respectively, with a 90\% uncertainty radius of $\rm 2.0^"$. This position was within the \swift-BAT error circle \citep{Swift:gcn25552}. Subsequently, the Ultra-Violet and Optical telescope (UVOT) onboard \swift, many ground-based optical and near-infrared telescopes began to follow-up observations of the GRB. We summarize these observations and also reduce data for the present analysis.
\subsubsection{Observations and data reduction}
The X-ray afterglow was monitored until $\rm \sim$ 7.8 $\times$\rm  10$^{6}$ $\rm s$ post-trigger beginning with window timing  (WT) mode. Finally, upper limits are obtained with PC mode data at $1.02 \times 10^7$ s ($\rm \sim 4~ months$). The XRT light curve and the spectrum has been obtained from the \swift online repository\footnote{\url{https://www.swift.ac.uk/}} hosted by the University of Leicester \citep{eva07,Evans:2009MNRAS}. The UVOT observed the source position 106 $\rm s$ after the BAT trigger
\citep{Swift:gcn25552}. An optical counterpart candidate consistent with the X-ray afterglow position 
had been discovered \citep{Swift_UVOT:gcn25570}. We obtained UVOT data from the \swift archive page \footnote{\url{http://swift.gsfc.nasa.gov/docs/swift/archive/}}. For the UVOT data reduction, we used \sw{HEASOFT} software version 6.25 with latest calibration database. We performed the reduction of the UVOT data using {\em uvotproduct} pipeline.
A source circular region of $\rm 5^"$ and a background region of $\rm 25^{"}$ aperture radius were extracted for the analysis. All the magnitudes have been converted to flux density using the UVOT zero point flux\footnote{\url{http://svo2.cab.inta-csic.es/theory/fps3/index.php?mode=voservice}}. Correction for Galactic and host galaxy extinction is not applied.
Results are plotted in Figure \ref{fig:multiwavelength_lc}(b).

 
 An evolving optical counterpart with preliminary magnitude m(r) $\rm \sim$ 16.0 was reported by \cite{Dabancheng:gcn25555} using the Half Meter Telescope (HMT-0.5m).
 \cite{GTC:gcn25565} observed the optical afterglow using 10.4 $\rm m$ GTC telescope. A red continuum with the Ca, H and K doublet absorption line was detected in the afterglow spectrum along with the emission lines of the SDSS galaxy (J025810.28-085719.2) at redshift $z$ = 0.0785. Other ground based optical and NIR telescopes also observed this evolving source in various filters
 \citep{MASTER:gcn25558, GCN:25560, NOT:gcn25563, GROND:gcn25569, KAIT:gcn25580, MMT:gcn25583, UKIRT:gcn25584, Liverpool:gcn25585, TNG:gcn25591, Liverpool:gcn25592, Liverpool:gcn25597, LCO:gcn25641, MPC:gcn25667}. We applied the Galactic extinction correction from the observed magnitudes using \cite{2011ApJ...737..103S}.
 An associated supernova also has been reported 
 by using photometry and spectroscopy observations \citep{Liverpool:gcn25623,  GROND:gcn25651,  MASTER:gcn25652, Liverpool:gcn25657,  Keck:gcn25664,  GTC:gcn25677, IKI:gcn25682}. 

The combined multiwavelength lightcurves are shown in Figure \ref{fig:multiwavelength_lc}. For completeness, we also showed low frequency data points \citep{uGMRT:gcn25627, MeerKAT:gcn25635,ATCA:gcn25676,NOEMA:gcn25589}. Through visual inspection, we can note that the optical flare is correlated with the X-ray flare.


\subsubsection{Analysis}\label{sec:XRT_analysis}

We have analysed the XRT spectrum in 0.3 - 10 $\rm keV$ band in \sw{XSPEC}. 
We used an absorption component along with the source spectral model. For this component, we chose a fixed Galactic column density of 5.60$\times$\rm 10$^{20}$\,$\rm cm^{-2}$, and a free intrinsic column density for the host redshift of 0.0785. We consider two models, a simple power-law model and a broken-power law model (\sw{bknpow}). We also searched for additional thermal and other possible components, however, such components are not present or not preferred by statistics. With each model, we included \sw{XSPEC} models \sw{phabs} and \sw{zphabs} for Galactic and intrinsic absorption, respectively. We also included \sw{redden} and \sw{zdust} model for interstellar extinction and reddening in the host, respectively. We considered MW, LMC and SMC extinction laws to get the reddening in the host.
All the parameters, along with various model, has been shown in Table \ref{tab:tr_spectral}.

We divide the XRT flux lightcurve into five phases (numbered I to V) based on its evolution. The initial emission in the X-rays shows different decay behavior in flux and flux density (@ 10 $\keV$). The flux decays with an index $\rm \sim$ 3 while the flux density (@ 10 $\keV$) shows a sporadically changing behavior
in the beginning. 
This is also reflected in our joint analysis of XRT and BAT data for phase I where we found that the spectrum could be described by a cutoff power-law model. The spectral index as shown in the lower panel  is also varying fast during phase I. A strong flare is also present in both  X-rays, \swift-UVOT and optical light curves beginning from $\sim 600$ $\s$ . We have modeled the X-rays in phase I by a power-law, III and IV by a power-law with a smooth break. The measured spectral and temporal parameters of the XRT light curve fitting are shown in Table \ref{tab:tr_spectral} \& \ref{tab:table_SED1}.

The external shock models predict certain closure relations between the spectral and temporal index in various regimes (cooling, density regimes, or an injection from the central engine).  These relations present tests without delving into details of the models ($e.g.,$ \citealt{Zhang:2004IJMPA, Gao2013NewAR}). Using the conventional notation,
$F_{\nu} \propto t^{-\alpha}\nu^{-\beta}$, we obtained $\alpha_X$, $\beta_X$ 
for \thisgrb afterglow in the X-ray bands. In Table \ref{tab:table_SED1}, we present the indices of the flux and flux density (@ 10 $\keV$) for the segments of the lightcurve. 
We particularly analyze segments III (flare) and IV (break) regions in detail, starting with phase IV.

\begin{figure*}
\centering
$~~~~~$\includegraphics[scale=0.23]{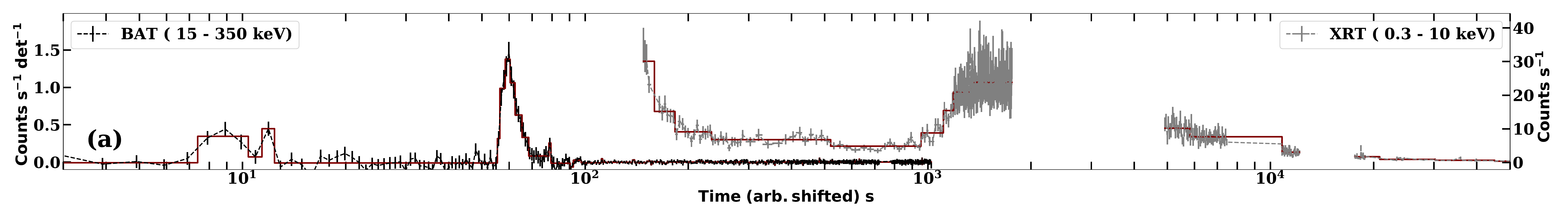}
\includegraphics[scale=0.23]{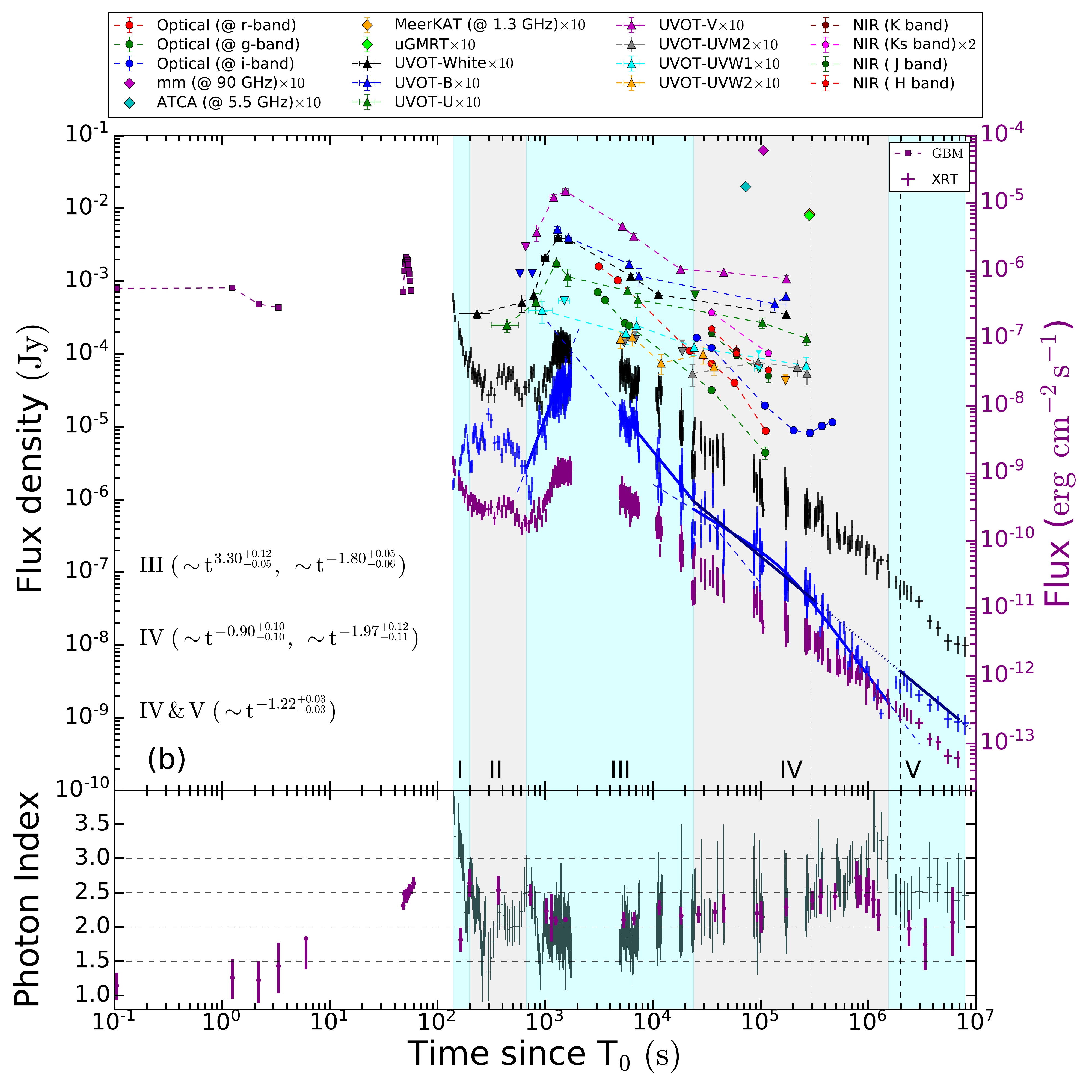}
\includegraphics[scale=0.22]{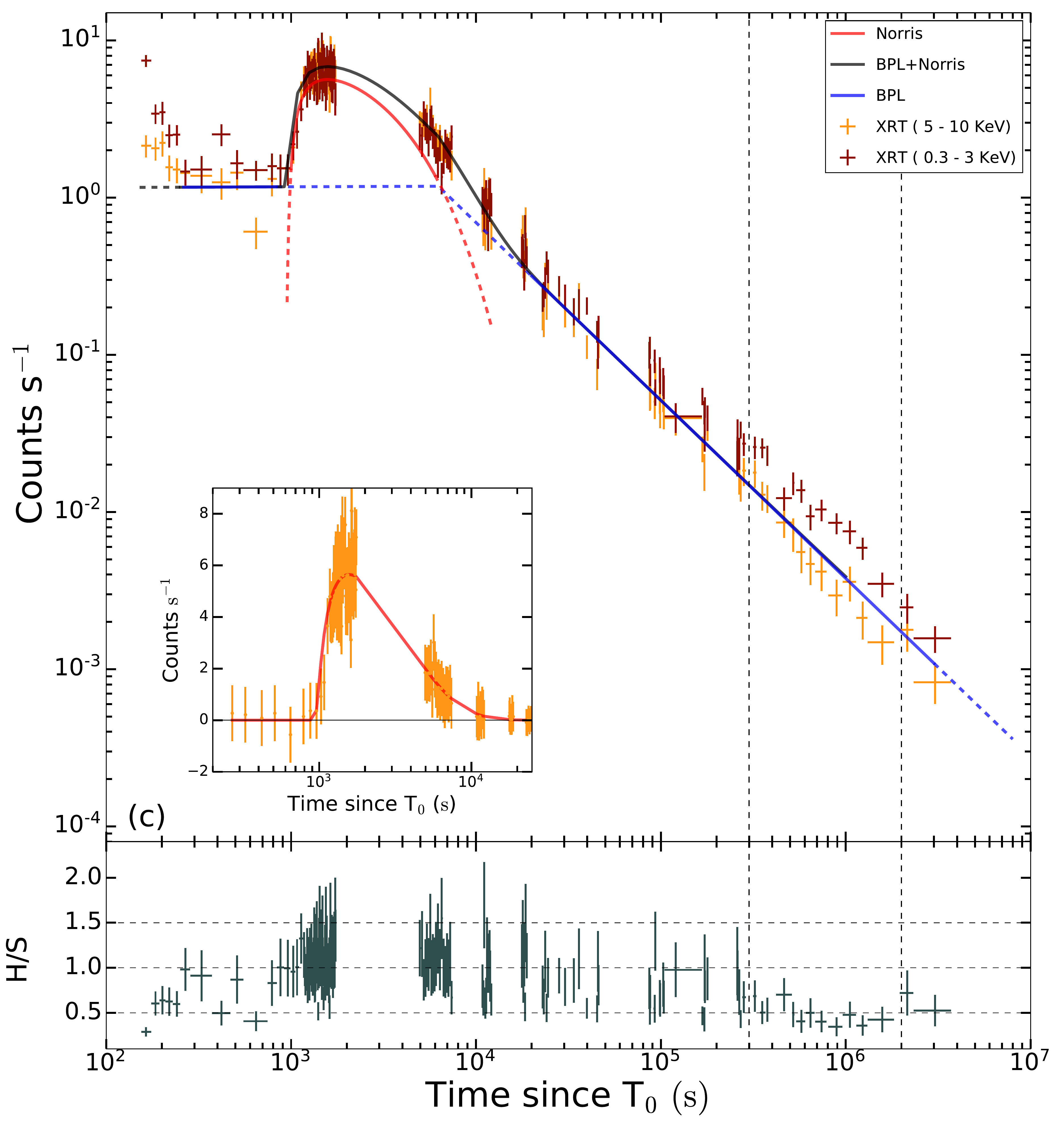}

 \caption{{\bf Multiwavelength light curves:} (a) BAT and XRT observation of all the episodes in log-linear scale, (b) {\it Upper Panel:} Flux ($\rm erg ~cm^{-2}~ s^{-1}$) for GBM (10 - 1000 $\keV$)  and XRT (0.3 -10 $\keV$) labeled on  right axis (purple axis), whereas the left axis shows the flux density ($\rm Jy$) in different wavelength regimes. The downside triangles denote the corresponding upper limits. For XRT, black data points are Flux density @ 1 keV and blue data points are flux density @ 10 keV. The shaded regions show the different episodes selected, and data between the vertical black-dashed lines is ignored for modeling is a possible rebrightening region. {\it Lower panel :} Photon index.  (c) XRT light curves: {\it Upper Panel :} XRT count rate in hard (5 - 10 $\keV$) and soft (0.3 - 3 $\keV$) band. The figure in the inset shows the background-subtracted flaring episode fitted well by the Norris model. {\it Lower panel :} The ratio of count rate in hard and soft bands (Hardness ratio). The vertical black dashed lines are the same as in (b). All XRT data are taken and reduced from the \swift online repository \citep{Evans:2009MNRAS}. 
 }
\label{fig:multiwavelength_lc}
\end{figure*}

The segment in the Flux lightcurve shows a shallow change ($0.3 \pm 0.3$) in the temporal decay index. Contrary to this, in the Flux density (@10 $\rm keV$), $\rm \alpha_X$ changes by $1.1 \pm 0.2$. Naively, one may tend to recognize the break with a jet break, however, if we carefully note, the photon index softens during $\sim 3\times10^{5}$ - $2\times 10^{6}$ s (vertical dashed lines in Figure \ref{fig:multiwavelength_lc} b). This is also reflected in the softening of the hardness ration (vertical dashed lines in Figure \ref{fig:multiwavelength_lc} c). Since after this period, the photon index settles down to its previous value before the spectral change, it suggests a rebrightening within the low energy Swift-XRT band. The passage of some spectral break frequency is also less likely due to the same reason as any frequency cross-over will cause an irreversible change in the spectral index. Phase IV and V (excluding region between the black dashed) is consistent with a typical decay with {$\alpha_X \sim {1.22\pm 0.03}$.} 
The UVOT data in this phase is possibly dominated by the contribution of the host. The observation in the i-band shows the rising part of a supernova \citep{Liverpool:gcn25623} contemporaneous with the break $\rm t_b$.

Considering adiabatic cooling without energy injection from the central engine. The inferred value of $\rm \alpha_X$ ($1.31 \pm 0.15$) from observed $\rm \beta_X$ ($1.21 \pm 0.1$) matches with the observed value ($1.22\pm0.03$) within errorbars for the spectral regime $\nu > \nu_c$ and ISM or Wind medium (Table \ref{tab:table_SED1}). We estimate the value of electron distribution index $p = 2.42 \pm 0.2$ from $p=2\beta$. 
We calculate the kinetic energy in the jet $\rm E_{K, iso} ~=~1.7\times{10^{52}}$ erg as according to Eq. 17 of \cite{Wang:2015ApJS} which is valid for the case $\nu_X > (\nu_m, \nu_c)$. We took a pre-break segment from phase IV (0.27 - 1.9 day) with a mean photon arrival time $\sim0.7$ day. The following values of the parameters are assumed: the fraction of post shock thermal energy in magnetic fields, $\rm \epsilon_B$ = 0.1 and in electrons, $\rm \epsilon_e$ = 0.1 and from the spectrum of this segment $p = 2.6$;
negligible inverse Compton scattering and typical value of ambient number density $\rm n$ = $1~ cm^{-3}$ \citep{Racusin09:jet, wang18:jet}. The efficiency ($\eta$) is then calculated by the ratio of the isotropic radiation emitted in the prompt emission and the total energy $E_{\gamma, iso}/(E_{\gamma, iso} + E_{K, iso}$). The efficiency $\eta$ is $\sim 1.3 \times 10^{-2}$ ($\sim$1.3\%). %

\subsection{X-ray flares}
Other than the two episodes detected in the prompt emission, we observed flaring activities in the X-ray afterglows. 
The average photon index in the non-flaring region scatters around the mean value near to the BAT photon index. For the flaring regions, e.g., the initial phase I, it has altogether different spectrum (CPL), and the same is reflected in a softer photon index. 
The softer and fast varying spectrum here is in support of the flaring activity. To uncover this, in the right panel, we have plotted count rate light curve in a low and high energy band optimized to see this effect, and we show that there is a flaring activity in the low energies. The spectra for the phases as well as Bayesian blocks obtained from the count rate light curve in 0.3 - 10 keV are fit, and the photon indices are plotted in the lower panel of Figure \ref{fig:multiwavelength_lc}(b). The photon index also scatters only about the mean values we have found for different phases. Some trending variations (the region between black-dashed lines) are as a result of possible low energy rebrightening.

Most prominent among the X-ray flares 
is phase III. The initial decaying phase (I) has a cutoff in the spectrum. The spectral index (plotted in the second panel of Figure \ref{fig:multiwavelength_lc}(b) varies fast during phases I and II and, therefore, are flare-like activities.  Episode III is a larger flare with a fast rise and decays with an index of $\rm \sim 2$. We discuss two scenarios for the origin of this flare: 

(a) {\it Reverse-shock emission in external shock}: The origin of flared emission can also be due to the reverse shock propagation into the ejecta medium \citep{Kobayashi:2007ApJ, Fraija:2017arXiv}. Since the X-ray flare is much delayed from the prompt phase, reverse shock occurs in a thin shell. The predicted temporal index for reverse shock SSC emission before the peak is $\alpha_X = 5(1-p)/4$ and after the peak is  $\alpha_X = (3p+1)/3$. Using the observed values for phase III, we estimate $p= 3.64_{-0.09}^{+0.04}$ and  $p= 1.47 \pm 0.05$ before and after the peak, respectively. Clearly, the reverse shock SSC emission is not a consistent interpretation. 

(b) {\it Late time central-engine activity:} Giant flares have been detected in X-rays and are associated to the central engine activity \citep{Falcone:2006ApJ, Dai:2012ApJ, Gibson:2017MNRAS}. These flares are superimposed on the underlying afterglow emission.
We have plotted count rate light curves for  hard (H: 5 - 10 $\keV$) and soft (S: 0.3 - 1.5 $\keV$) bands in Figure \ref{fig:multiwavelength_lc}(b). The hardness ratio (H/S) is shown in the lower panel of this figure which reflects the comparative strength of the signal in these two bands. This uncovers plateau phase before the flare in H-band. The peak rate of the flare beginning at $\rm \sim 1000 s$ is $\rm \sim 5.8$ times higher in comparison to the plateau phase. We model the overall light curve using a combination of broken powerlaw (BPL) for the underlying afterglow and Norris model \footnote{ $I(t) = A \lambda {\rm~exp}{[-\tau_1/(t - t_i) - (t - t_i)/\tau_2]}$, where $A$ is pulse amplitude, $\tau_1$, $\tau_2$ are rise and decay time of the pulse respectively and $\rm t_{i}$ is the start time. The fit parameters are $\tiny{A = 5.6 \pm 0.2}$, $\tiny{\tau_1 = 161_{-46}^{+61}}$, $\tiny{\tau_2 = 2742_{-280}^{+250}}$ and $\tiny{t_i = 909_{-41}^{+33}}$. The pulse width is $\rm w$ = $\tau_2(1+4\sqrt{\tau_1/\tau_2})^{0.5} $. The rise time $\rm t_{r} \sim 553$ s, decay time rise time is $\rm t_{d} \sim  3293$  s and peak time $\rm t_{pk}$ is $1573$ s.}
for the flare \citep{Norris:2005ApJ}. We calculate the pulse width ($\rm w$) and asymmetry of the pulse using the measured values of pulse rise and decay time and find that $\rm w$ is $\sim$3848 s and asymmetry $k$ using the relation ({$\rm k$ = $\tau_2/{\rm w}$}) is 0.71. The ratio of rise time ($\rm t_{r}$) to decay time ($\rm t_{d}$) ratio is $\sim 2$ is well within the $\rm t_{r}/\rm t_{d}$ distribution. The origin of the X-ray flares (bumps observed in X-ray emission) in GRBs is widely discussed and studied \citep{Ioka:2005ApJ, Chincarini:2010MNRAS, Curran:2008AandA}. Among the X-ray flares observed, the late time flares (with $\rm t_{pk} \sim 1000$ s are also specifically studied \citep{Curran:2008AandA, Bernardini:2011AandA}. The  relative varibility defined as $\rm w$/$\rm t_{pk}$ is $\sim 2.4$. This implies the kinematically allowed regions in the "Ioka plot" for the X-ray flare in \thisgrb are where the emission is originating from refreshed shocks or patchy shells, and does not satisfy the general requirement ($\rm w$/$\rm t_{pk} < 1$) for an internal activity \citep{Ioka:2005ApJ, Curran:2008AandA, Chincarini:2010MNRAS, Bernardini:2015MNRAS}. 
The isotropic X-ray energy $\rm E_{X,iso}$ of the flaring phase to be $\rm 4.49 \times 10^{49}~ erg$.

\section{Summary and Discussion}
In this letter, we highlighted the unusual spectral features of episodic activities in \thisgrb from the prompt emission to afterglows. We found that Amati and Yonetoku relations are satisfied for GRBs with multiple episodes separated by quiescent phases. But, the first episode of \thisgrb is the only outlier. It also does not satisfy the pulse-wise Amati correlation known for well-separated individual pulses of GRBs \citep{Basak:2013MNRAS}. In the hardness-duration diagram and spectral-lags, this GRB emission is consistent with long GRBs \citep{Bhat:2016ApJS, Goldstein:2017ApJ}. 

\emph{\bf Shock breakout origin of the prompt emission?:
Some of the low-luminosity GRBs are not compatible with Amati correlation (see Figure \ref{fig:Amati_Yonetoku}). 
The radiation in LLGRBs is powered by shock breakout and the energetic satisfy a fundamental correlation
$T_{90}\sim 20~{\rm s}~(1+z)^{-1.68}\left(\frac{E_{\rm \gamma,iso}}{10^{46}~{\rm erg}}
\right)^{1/2}\left(\frac{E_{\rm p}}{50~{\rm keV}}\right)^{-2.68}$ \citep{Nakar:2012ApJ}. 
For the parameters of the first episode which lies outside in the Amati plane ($3 \sigma$), $\rm E_{\gamma, iso}\sim 3.2 \times 10^{49}$ erg and $\rm E_p \sim 120$ keV, and z = 0.0785, the predicted shock break-out duration is $\sim 9.5$ s which is similar to \tninty or the duration of the episode taken for spectral analysis ($\sim 8.7$ s). This is a favorable evidence for the shock breakout interpretation of this episode. For the second episode considering $\rm E_p$ of 10 keV, the predicted $T_{90}$ is $>$ 18135 s. This is much larger than the observed value of 10.4 s. 
Secondly, for a shock break out, we do not expect the variability on a scale much shorter than the typical timescale of the pulse (e.g. the rise time $\sim1$ s in case of \thisgrb). From the Bayesian blocks \citep{scargle:2013} constructed for the GBM-NaI \& the BAT lightcurves, we found that no short time variability is found (with a false alarm probability
$\rm p_0$ = 0.35). The observed prompt emission efficiency is $\eta \sim 1.3\%$ which is normal as observed for long and short GRBs \citep{Zhang:2007ApJ, Racusin:2011ApJ} and not low ($\sim 10^{-4}$) as observed for shock breakout and LLGRBs \citep{Gottlieb:2018MNRAS}. 
The soft episode is separated from the low luminosity hard episode by a quiescent phase, and has a spectrum which is a power-law with value typically observed for a Band high energy power-law spectrum in GRBs. This is contrary to the soft thermal emission with no significant gap expected in shock breakout model. 
A shock break out interpretation would also raise the upper limit of the shock breakout luminosity $\sim$10 times of the previous limit of $10^{48}$ erg $\rm s^{-1}$ \citep{Zhang:2012ApJ756} and critical limit set in \citealt{Matsumoto:2020MNRAS}. Hence, our analysis suggests that emission in \thisgrb is not caused by a shock breakout}.

{\bf The late time flare observed is unusual as the relative variability is atypical and the flare is also observed simultaneously in optical bands. A similar example with $\rm w/t_{pk} > 1$ is GRB 050724 which was extensively studied. Detailed studies showed that the flare in GRB 050724 could be interpreted within different frameworks \citep{Panaitescu:2006MNRAS, Malesani:2007AA, Lazzati:2007MNRAS, Bernardini:2011AandA}. Another possibility for the origin of the flare includes fallback accretion on a newborn magnetar \citep{Gibson:2017MNRAS}. 

In late time LAT emission, there are some photons ($>$ 100 MeV) associated with the source, which may have originated in synchrotron emission as GRB 190114C and the HESS detection, which lies above the maximum synchrotron energy similarly might have inverse Compton emission. The X-ray observations during the flare emission, which occurs after 1000 s, cannot be explained by the reverse shock SSC temporal relations \citep{Kobayashi:2007ApJ, Fraija:2017arXiv}. 
An excess emission in low energy X-ray band (0.3 - 3 keV) is seen at times $\sim 3\times 10^{5} - 2\times 10^{6}$ s  and softening is observed in hardness ratio during this.
The time-averaged $\gamma$-ray luminosity for \thisgrb is one order of magnitude above the threshold for internal engine activity ($\sim 10^{48}$ $\rm erg ~s^{-1}$), which disfavours shock-breakout origin \citep{Zhang:2012ApJ756}}. 
Given the detection in the TeV band, it is likely that the viewing angle is closer to the jet axis, as larger viewing angles may not provide sufficient Doppler boosting. This implies the faintness of GRB episodes is intrinsic rather than the effect of the viewing angle. The early signature of an emerging supernova emission in optical i-band, as shown in Figure \ref{fig:multiwavelength_lc}(a), further supports this hypothesis. However, a deeper understanding 
would require incorporating detailed modelling of the source, including the HESS observation and the study of the associated supernova. 


\section*{Acknowledgments}
We thank Dr. K. L. Page and Dr. Phil Evans of {\it Swift helpdesk} for helpful discussions to deal with {\it Swift}-XRT data. We thank Prof. Bing Zhang for a discussion on the possible interpretation of the results. We also thank Prof. A.R. Rao, Dr. Gor Oganesyan, Dr. A. Tsvetkova and Dr. N. Fraija for fruitful discussions.  We also acknowledge and are thankful for the constructive comments by the anonymous referee.
This work is supported by National Key Research and Development Programs of China (2018YFA0404204) and The National Natural Science Foundation of China (Grant Nos. 11833003) and The Innovative and Entrepreneurial Talent Program in Jiangsu, China. BBZ acknowledges support from a national program for young scholars in China.
RG and SBP acknowledge BRICS grant {DST/IMRCD/BRICS/PilotCall1/ProFCheap/2017(G)}. PSP
acknowledges SYSU-Postdoctoral Fellowship. RB acknowledges funding from the European Union’s Horizon 2020 research and innovation programme under the Marie Sklodowska-Curie grant agreement n. 664931. This research has made use of data obtained through the HEASARC Online Service,
provided by the NASA-GSFC, in support of NASA High
Energy Astrophysics Program.


\input{table_tr.tex}

\input{table_SED.tex}
\bibliography{GRB190829A}
\bibliographystyle{aasjournal}

\clearpage
\begin{appendix}

\label{appendix}
\section{Two phase GRBs with redshift detection}
\label{sec:two_phases}
\begin{figure}[h]
\centering
  \includegraphics[scale=0.38]{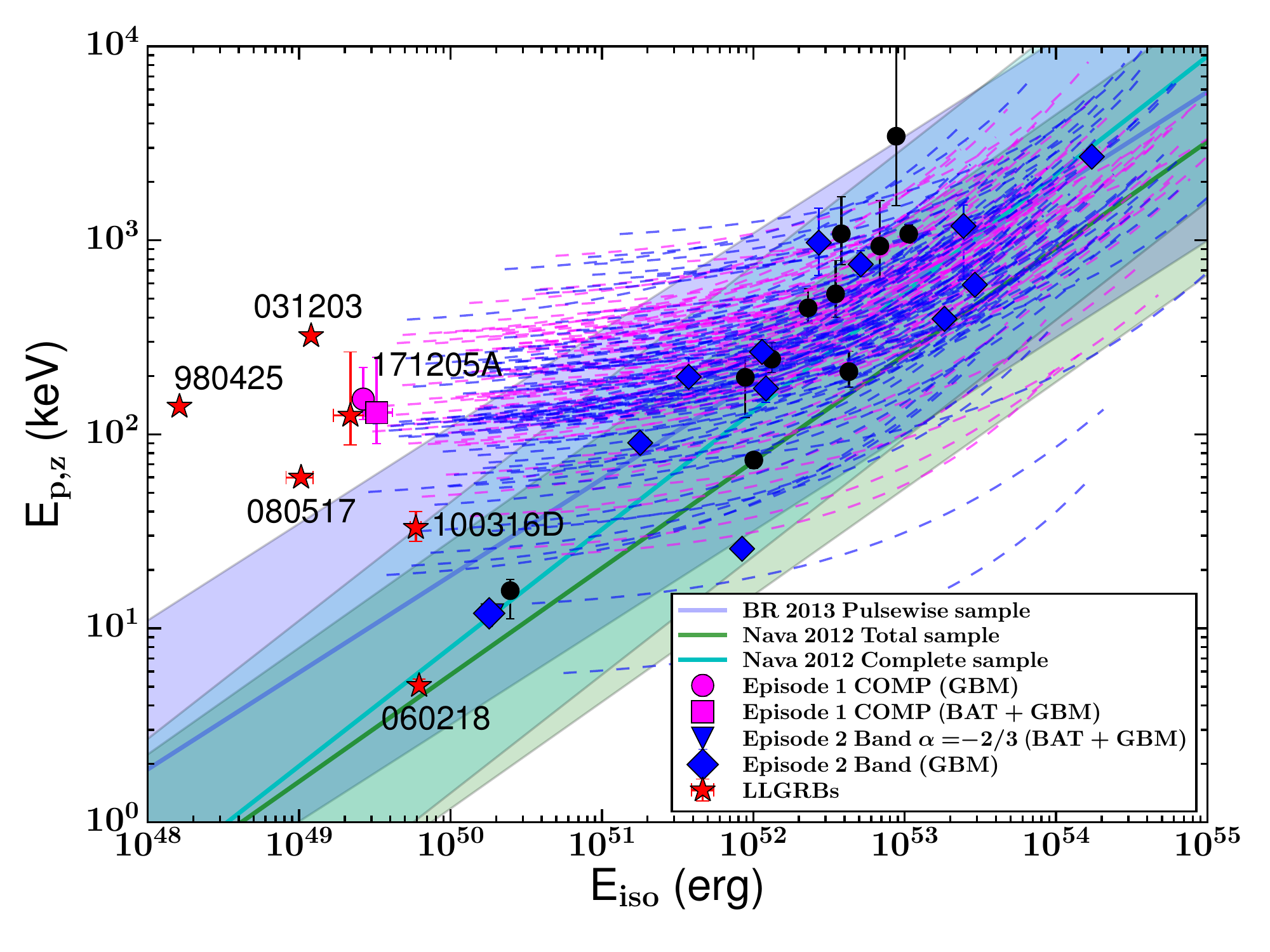}
  \includegraphics[scale=0.38]{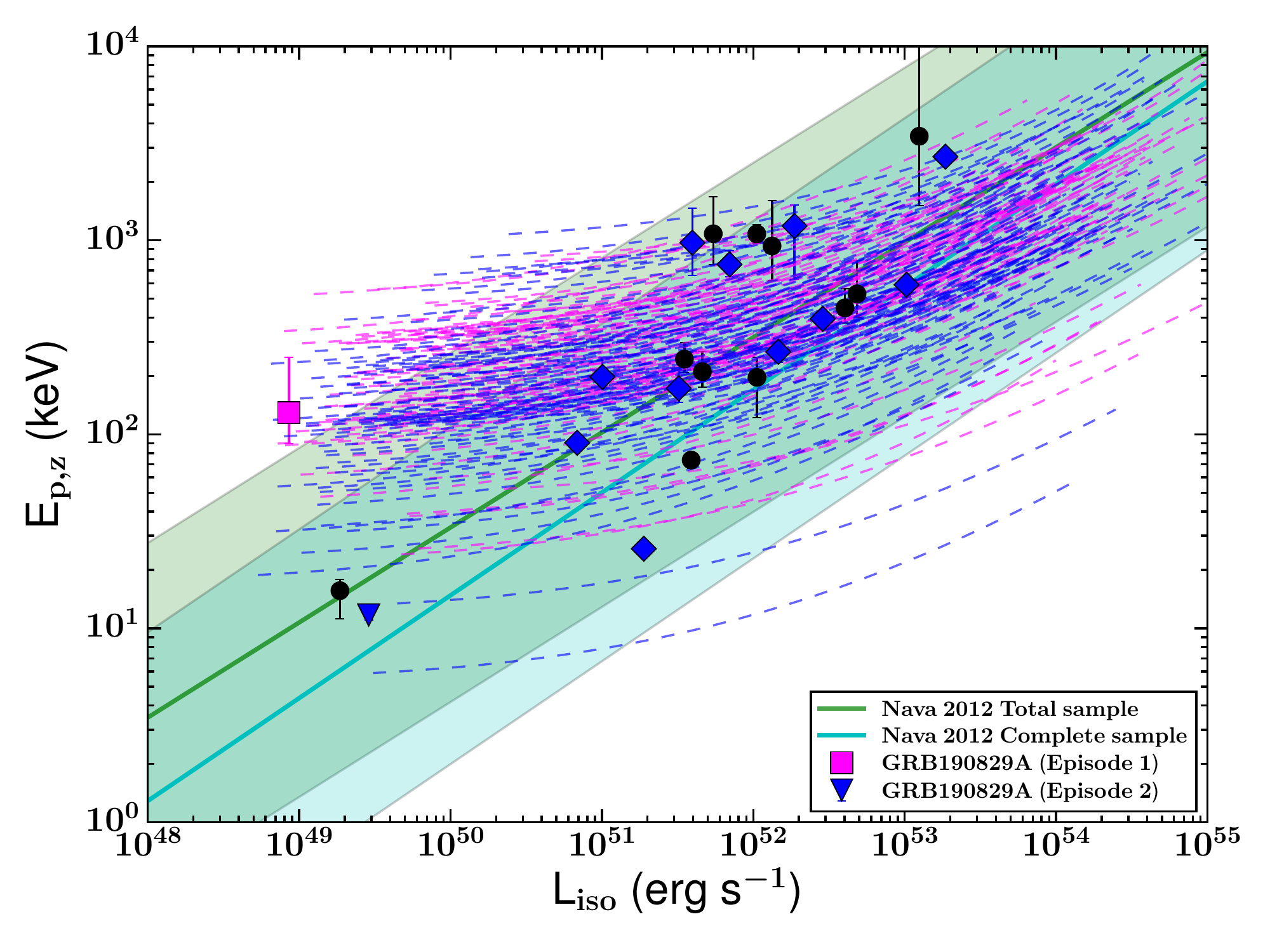}
\caption{ {\bf Correlations for two episode GRBs:} Same as Figure \ref{fig:Amati_Yonetoku}. Tracks for all the GRBs without redshift are shown. BAT + GBM data-point is shown for episode 1. For episode 2, the low energy index  ($\alpha = -2/3$) is set to the slow cooling limit.}
\label{fig:Amati_Yonetoku_full}
\end{figure}

For the sample of GRBs with two episodes as reported in \cite{Lan:2018ApJ}, 11 have measured redshift
\footnote{\url{http://www.mpe.mpg.de/~jcg/grbgen.html}}. We extracted the spectrum using the same criteria as described in Section \ref{sec:prompt_emission} and performed time integrated analysis for each episode. We fit \sw{COMP} and Band models to the background subtracted spectral data and compared the models using BIC.
For studying correlations, we calculated the flux within the energy range specified by 1/1+z $\rm keV$ to 10/1+z $\rm MeV$  and computed the $\rm E_{\rm iso}$ and $\rm L_{iso}$. For other GRBs in the sample, spectral properties of each episodes are well constrained but there is no redshift estimate. Here we calculated the $\rm E_{\rm iso}$ and $\rm L_{iso}$ values by varying
redshift ranging from 0.01 to 10 shown by continuous tracks in Figure \ref{fig:Amati_Yonetoku_full}. We assume following cosmology parameters Hubble parameter, $\rm H_{0}$ = 71 $\rm km$ $\rm s^{-1}$ $\rm Mpc^{-1}$, total matter density, $\rm \Omega_{M}$ = 0.27, and  dark energy density, $\rm \Omega_{\Lambda}$ = 0.73. 

\input{table_two_episodes.tex}
\input{table_UVOT.tex}

\end{appendix}

\end{document}

%% file: table_tr.tex
\begin{table*}[!h]
\caption{Time-resolved spectral fitting of prompt emission Episode 1 \& Episode 2 (\fermi-GBM) and spectral fitting of the afterglow phases.}
\setlength{\tabcolsep}{2pt}
\flushleft
\begin{center}
\footnotesize
    Prompt Emission Episode 1
\end{center}
\begin{tiny}
\setlength\tabcolsep{1.0pt} 
\begin{tabular}{|p{1.45cm}|p{1.7cm}p{1.7cm}|p{1.7cm}p{1.7cm}p{1.7cm}|p{1.7cm}p{1.72cm}p{1.75cm}p{1.75cm}|p{1.75cm}p{1.75cm}p{1.75cm}|}
\cline{1-10}
\cline{1-10}
 \multicolumn{1}{|c|}{Time (s)} & \multicolumn{2}{c|}{PL} & \multicolumn{3}{c|}{COMP} & \multicolumn{4}{c|}{Band}  &\\
\cline{2-10}
 $\rm $ & $\rm \alpha$ & \sw{pgstat}(BIC) & $\rm \alpha$ & $\rm E_{p}$ $\rm (keV)$ & \sw{pgstat}(BIC) & $\rm \alpha$ & $\rm \beta$ & $\rm E_{p}$ $\rm (keV)$ & \sw{pgstat}(BIC) \\ 
  &  & dof = $\rm 467$ &  &  &  dof = $\rm 466$ &  &  &  & dof = $\rm 465$ \\ 
 \cline{1-10}
 -0.64 - 0.85 & $1.49_{-0.05}^{+0.06}$ & $463.6(494.4)$ & $1.14_{-0.21}^{+0.19}$    & $311_{-112}^{+324}$ & $453.5(490.5)$ & $-1.01_{-0.28}^{+0.36}$ & $-1.89_{+1.89}^{+0.26}$ & $194_{-84}^{+314}$ & $452.3(495.4)$ \\
 0.85 - 1.64  & $1.63_{-0.07}^{+0.08}$ & $490.8(521.6)$ & $1.26_{-0.31}^{+0.27}$    & $155_{-52}^{+279}$  & $484.2(521.1)$ & $-0.75_{-0.64}^{+0.75}$ & $-1.88_{-0.95}^{+0.19}$ & $65_{-35}^{+132}$  & $480.8(523.9)$ \\
 1.64 - 2.72  & $1.78_{-0.08}^{+0.08}$ & $532.9(563.7)$ & $1.22_{-0.33}^{+0.28}$    & $86_{-19}^{+42}$    & $517.1(554.0)$ & $-1.07_{-0.32}^{+0.72}$ & $-2.46_{+2.46}^{+0.45}$ & $71_{-30}^{+38}$   & $516.3(559.4)$ \\
 2.72 - 3.96  & $1.87_{-0.09}^{+0.10}$ & $480.0(510.8)$ & $1.43_{-0.40}^{+0.34}$    & $68_{-18}^{+66}$    & $474.6(511.6)$ & $-1.20_{-0.46}^{+0.66}$ & $-2.30_{+2.30}^{+0.32}$ & $52_{-17}^{+41}$   & $472.8(515.9)$ \\
 3.96 - 8.06  & $2.18_{-0.12}^{+0.14}$ & $540.8(571.6)$ & $1.83_{-0.45}^{+\infty}$  & $21_{-16}^{+18}$    & $537.2(574.2)$ & $-0.43_{-1.99}^{+0.43}$ & $-2.34_{+2.34}^{+0.20}$ & $19_{-6}^{+18}$    & $536.5(579.6)$ \\
 \cline{1-10}
\end{tabular}
\end{tiny}

\begin{center}
\footnotesize
    Prompt Emission Episode 2
\end{center}
\begin{tiny}

\begin{tabular}{|l|ll|lll|llll|lll|}
\hline

\multicolumn{1}{|c|}{Time (s)} &\multicolumn{2}{c|}{PL} & \multicolumn{3}{c|}{CPL} & \multicolumn{4}{c|}{Band} & \multicolumn{3}{c|}{BB+PL}\\
\cline{2-13}
& $\rm \alpha$ & \sw{pgstat}(BIC) & $\rm \alpha$ & $\rm E_{c}$ $\rm (keV)$ & \sw{pgstat}(BIC) & $\rm \alpha$ & $\rm \beta$ & $\rm E_{ p}$ $\rm (keV)$ & \sw{pgstat}(BIC) & $\rm \alpha$ & kT $\rm (keV)$ & \sw{pgstat}(BIC)\\ 
  &  & dof= $\rm 450$ &  &  & dof= $\rm 449$ &  &  &  & dof = $\rm 448$ & & & dof = $\rm 448$\\
\hline

47.04 - 49.03 & $2.31_{-0.06}^{+0.06}$ & $452.4(483.0)$ & $2.12_{-0.15}^{+0.14}$ & $240_{-115}^{+605}        	  $  & $446.1(482.8)$ & $  0.75_{-0.95}^{+\infty}  $ & $ -2.38_{-0.06}^{+0.05} $ & $ 14.02_{-0.18}^{+1.44}   $  & $442.6(485.5) $ & $ 2.20_{-0.12}^{+0.11} $ & $ 4.2_{-0.8}^{+1.7}    $ & $ 446.9(489.7) $\\
49.03 - 49.68 & $2.48_{-0.06}^{+0.06}$ & $471.3(501.9)$ & $2.27_{-0.17}^{+0.18}$ & $182_{-85}^{+745}      $  & $466.3(503.0)$ & $ -1.00_{-1.12}^{+1.00}    $ & $ -2.57_{-0.25}^{+0.08} $ & $ 12.29_{-5.17}^{+2.18}   $  & $462.0(504.8) $ & $ 2.43_{-0.12}^{+0.12} $ & $ 5.0_{-1.3}^{+4.2}    $ & $ 464.6(507.4) $\\
49.68 - 50.21 & $2.43_{-0.06}^{+0.06}$ & $573.2(603.8)$ & $2.41_{-0.13}^{+0.07}$ & $2278_{-1967}^{+\infty}      $  & $573.1(609.8)$ & $  0.98_{-0.98}^{+\infty}  $ & $ -2.44_{-0.06}^{+0.06} $ & $ 10.49_{-0.11}^{+1.74}   $  & $571.5(614.4) $ & $ 2.39_{-0.13}^{+0.15} $ & $ 4.4_{-4.4}^{+5.6}    $ & $ 572.0(614.9) $\\
50.21 - 50.66 & $2.46_{-0.05}^{+0.06}$ & $400.1(430.6)$ & $2.33_{-0.16}^{+0.14}$ & $328_{-185}^{+6799}          $  & $397.0(433.6)$ & $  1.00_{-1.00}^{+\infty}  $ & $ -2.51_{-0.07}^{+0.07} $ & $ 11.94_{-0.17}^{+1.42}   $  & $392.4(435.3) $ & $ 2.38_{-0.12}^{+0.15} $ & $ 3.4_{-2.4}^{+\infty} $ & $ 398.0(440.8) $\\
50.66 - 51.11 & $2.50_{-0.06}^{+0.06}$ & $501.3(531.9)$ & $2.50_{-0.08}^{+0.04}$ & ---    & $501.3(538.1)$ & $  0.96_{-0.96}^{+\infty}  $ & $ -2.51_{-0.04}^{+0.04} $ & $  9.44_{-0.10}^{+1.99}   $  & $501.0(543.6) $ & $ 2.34_{-0.14}^{+0.14} $ & $ 2.9_{-0.8}^{+0.6}    $ & $ 497.0(539.5) $\\
51.11 - 51.52 & $2.40_{-0.05}^{+0.06}$ & $488.0(518.2)$ & $2.26_{-0.16}^{+0.14}$ & $299_{-164}^{+3034}          $  & $484.0(520.8)$ & $ -1.36_{-0.50}^{+1.36}    $ & $ -2.52_{-0.11}^{+0.09} $ & $ 13.41_{-8.05}^{+3.06}   $  & $475.2(518.0) $ & $ 2.30_{-0.12}^{+0.10} $ & $ 4.8_{-0.8}^{+1.6}    $ & $ 475.0(517.9) $\\
51.52 - 51.93 & $2.44_{-0.05}^{+0.06}$ & $497.8(528.4)$ & $2.40_{-0.13}^{+0.09}$ & $1324_{-1048}^{+\infty}      $  & $497.4(534.1)$ & $ -1.81_{-0.65}^{+1.81}    $ & $ -2.49_{-0.10}^{+0.08} $ & $  5.92_{-2.46}^{+7.44}   $  & $495.0(537.8) $ & $ 2.34_{-0.13}^{+0.11} $ & $ 4.1_{-0.8}^{+1.7}    $ & $ 493.0(535.8) $\\
51.93 - 52.37 & $2.44_{-0.06}^{+0.05}$ & $515.1(545.7)$ & $2.44_{-0.10}^{+0.04}$ & ---   & $515.1(551.8)$ & $  0.98_{-0.80}^{+\infty}  $ & $ -2.48_{-0.07}^{+0.07} $ & $ 12.07_{-0.13}^{+1.48}   $  & $508.4(551.2) $ & $ 2.22_{-0.15}^{+0.13} $ & $ 3.7_{-0.4}^{+0.5}    $ & $ 503.0(545.8) $\\
52.37 - 52.80 & $2.47_{-0.06}^{+0.06}$ & $431.6(462.2)$ & $2.34_{-0.16}^{+0.14}$ & $315_{-174}^{+5454}       	  $  & $428.5(465.3)$ & $  0.91_{-1.49}^{+\infty}  $ & $ -2.52_{-0.05}^{+0.05} $ & $ 12.05_{-0.13}^{+1.50}   $  & $425.9(468.7) $ & $ 2.40_{-0.13}^{+0.06} $ & $ 4.1_{-1.0}^{+3.9}    $ & $ 428.3(471.1) $\\
52.80 - 53.26 & $2.46_{-0.06}^{+0.06}$ & $501.8(532.4)$ & $2.36_{-0.15}^{+0.14}$ & $393_{-237}^{+\infty}        $  & $500.0(536.7)$ & $  1.00_{-10.95}^{+\infty} $ & $ -2.52_{-0.08}^{+0.07} $ & $ 12.37_{-0.14}^{+1.49}   $  & $494.8(537.7) $ & $ 2.32_{-0.13}^{+0.13} $ & $ 3.6_{-0.6}^{+1.0}    $ & $ 496.4(539.2) $\\
53.26 - 53.77 & $2.47_{-0.05}^{+0.06}$ & $453.8(484.4)$ & $2.39_{-0.13}^{+0.11}$ & $568_{-366}^{+\infty}        $  & $452.1(488.8)$ & $ -2.05_{-\infty}^{+2.05}  $ & $ -2.50_{-0.08}^{+0.07} $ & $  10.39_{-\infty}^{+2.33} $  & $454.0(496.6) $ & $ 2.42_{-0.12}^{+0.17} $ & $ 4.0_{-4.0}^{+\infty}  $ & $ 452.6(495.5) $\\
53.77 - 54.39 & $2.53_{-0.06}^{+0.06}$ & $481.2(511.8)$ & $2.47_{-0.16}^{+0.11}$ & $682_{-503}^{+\infty}        $  & $481.0(517.5)$ & $ -1.63_{-0.87}^{+1.63}    $ & $ -2.63_{-0.11}^{+0.09} $ & $  8.16_{-5.08}^{+5.40}   $  & $474.0(516.8) $ & $ 2.38_{-0.15}^{+0.13} $ & $ 4.0_{-0.5}^{+1.0}    $ & $ 470.6(513.4) $\\
54.39 - 55.08 & $2.56_{-0.06}^{+0.07}$ & $419.0(448.7)$ & $2.52_{-0.19}^{+0.11}$ & $739_{-592}^{+\infty}        $  & $417.7(454.4)$ & $ -1.13_{-2.15}^{+1.13}    $ & $ -2.64_{-0.15}^{+0.09} $ & $ 10.00_{-4.19}^{+2.96}   $  & $413.7(456.5) $ & $ 2.47_{-0.17}^{+0.13} $ & $ 4.0_{-0.7}^{+1.9}    $ & $ 412.3(455.2) $\\
55.08 - 56.02 & $2.55_{-0.07}^{+0.06}$ & $517.2(547.8)$ & $2.53_{-0.17}^{+0.06}$ & $3607_{-3402}^{+\infty}      $  & $517.2(553.9)$ & $  0.99_{-0.79}^{+\infty}  $ & $ -2.60_{-0.09}^{+0.08} $ & $ 11.53_{-0.12}^{+1.38}   $  & $511.4(554.3) $ & $ 2.41_{-0.17}^{+0.14} $ & $ 3.5_{-0.6}^{+1.4}    $ & $ 513.0(554.9) $\\
56.02 - 57.50 & $2.55_{-0.07}^{+0.07}$ & $498.3(528.9)$ & $2.55_{-0.12}^{+0.07}$ & ---  & $498.3(535.0)$ & $ -0.40_{-3.13}^{+0.40}    $ & $ -2.61_{-0.10}^{+0.08} $ & $ 10.71_{-2.81}^{+1.85}   $  & $494.4(537.2) $ & $ 2.29_{-0.17}^{+0.17} $ & $ 3.3_{-0.4}^{+0.4}    $ & $ 487.4(530.2) $\\
57.50 - 62.46 & $2.64_{-0.08}^{+0.09}$ & $507.1(537.7)$ & $2.64_{-0.10}^{+0.08}$ & --- & $507.1(543.8)$ & $  0.95_{-3.44}^{+\infty} $ & $-2.70_{-0.08}^{+0.07}$ & $10.44_{-0.12}^{+1.45}$  & $504.0(546.8)$ & $2.16_{-0.21}^{+0.21}$ & $2.9_{-0.3}^{+0.2}$ & $488.3(531.1)$ \\ \hline
\end{tabular}
\end{tiny}

\begin{center}
\footnotesize
    Afterglow
\end{center}
\begin{tiny}
\begin{tabular}{|p{3.0cm}|p{1.65cm}|p{4.468cm}p{4.468cm}p{4.05cm}|}
\hline
\cline{1-5}
\multicolumn{1}{|c|}{(Phase) Time (s) }&
\multicolumn{1}{c|}{Detectors}&
\multicolumn{3}{c|}{\sw{Tbabs*zTBabs*cflux}*PL}  \\
\cline{3-5}
  & & $n_{H}$ &$\rm \alpha$ & \sw{Statistics}(BIC) \\
\cline{1-5}
(I) 104-228 & XRT +BAT & $0.84_{-0.20}^{+0.27}$ & $2.11_{-0.17}^{+0.20}$& 125.1/103  \\
(IV-pre-break)  23.7k - 167k & XRT   &  $1.03_{-0.12}^{+0.13}$ &  $2.02_{-0.11}^{+0.12}$ &532.1/541   \\
(IV post-break) 167k - 1550k & XRT   &  $1.12_{-0.12}^{+0.13}$ &  $2.35_{-0.13}^{+0.13}$ &  486.76/489      \\
\hline
\end{tabular}
\begin{tabular}{|p{3.0cm}|p{1.65cm}|llll|lllllp{1.1cm}|}
\hline
 \multicolumn{1}{|c|}{(Phase) Time (s) } &
 \multicolumn{1}{c|}{Detectors}&
 \multicolumn{4}{c|}{\sw{redden*phabs*zdust}*PL}& \multicolumn{6}{c|}{\sw{redden*phabs*zdust}*\sw{bknpower}} \\
\cline{3-12}
& & $ \rm n_{H}$ &$\rm \Gamma$ & \sw{Statistics}(BIC) & $\rm E_{B-V}$ & $n_{H}$ & $\rm \Gamma_1$ & $\rm \Gamma_2$ & $\rm E_b$ & \sw{Statistics}(BIC) & $\rm E_{B-V}$  \\
\cline{1-12}
 
(III-pre) 147 - 620  & UVOT + XRT     & $0.86_{-0.20}^{+0.23}$ & $1.84_{-0.15}^{+0.14}$& 39.7/26 &$1.14_{-0.28}^{+0.25}$ 
& $0.80_{-0.38}^{+0.32}$ & $1.50_{-0.54}^{+0.29}$  & $3.14_{-0.44}^{+0.85}$ & $3.54_{-0.89}^{+3.54}$& 25.9/24& $0.67_{-0.63}^{+0.43}$  \\
(III-rise)  620 - 1120 & UVOT + XRT     & $1.26_{-0.23}^{+0.33}$ & $1.92_{-0.12}^{+0.21}$ &56.5/30&$1.08_{-0.25}^{+0.48}$
-- & -- &    --    & -- & --  & -- & --            \\
(III-top)  1120-1700  & UVOT + XRT   & $1.41_{-0.07}^{+0.07}$ & $1.99_{-0.04}^{+0.05}$ &404.3/353&$1.33_{-0.10}^{+0.11}$     & $1.36_{-0.07}^{+0.07}$ & $1.84_{-0.06}^{+0.07}$  & $2.59_{-0.55}^{+0.81}$& $2.15_{-0.09}^{+0.17}$  &378.9/351  & $1.04_{-0.12}^{+0.14}$  \\
\cline{1-12}
\end{tabular}
$a$: $\chi^2$/dof for joint BAT-XRT data and UVOT-XRT data and \sw{C-stat} for XRT data. All errors are quoted at $1 \sigma$ (nominal 68\%) confindence level.
 $\rm N_{H} (z)$ : in units of $\rm 10^{22}$ $\rm cm^{-2}$;
$\rm  E_{b}$ are the break and cutoff energies, respectively in $\rm keV$. Reddening parameter in host is obtained for SMC extinction law. 
\end{tiny}
\label{tab:tr_spectral}
\end{table*}

%% file: table_SED.tex
\begin{table*}
\begin{center}
\renewcommand{\arraystretch}{1}\addtolength{\tabcolsep}{3pt}
\caption{The best-fit values of spectral and temporal indices for episodes marked in Figure \ref{fig:multiwavelength_lc}.}
\begin{tabular}{ |c | c  c c c | c c c c|}
 \hline 
\tiny{X-rays} &  \tiny{Intervals ($\rm s$) from}  & \tiny{Index}    & \tiny{ -$\rm log(L)/(n, k)$} & \tiny{Index}  & \tiny{Intervals ($\rm s$) from} & \tiny{Index} &\tiny{ -$\rm log(L)/(n, k)$} & \tiny{Index}\\ &
\tiny{Flux ($\rm erg ~cm^2 ~s^{-1}$)} & \tiny{$\rm \alpha$}  &  & \tiny{$\rm \beta$} & \tiny{Flux $\rm @$ 10 $\rm keV$ ($\rm Jy$ $@ 10 ~ \rm keV$)} & \tiny{($\rm \alpha_X$)}   &\tiny{}  & \tiny{($\rm \beta_X$)} \\ 
\hline
\tiny{I}   	        & \tiny{89.54 - 147.93}  & \tiny{$2.99_{-0.30}^{+0.20}$}  	&  \tiny{362/(16, 3)} &\tiny{$1.18_{-0.24}^{+0.26}$} & --- &--- & --- & ---\\	

\tiny{II}   	        & \tiny{147.93 - 619.84}  & \tiny{---} 	&  \tiny{---} & \tiny{$1.46_{-0.22}^{+0.23}$} & --- & --- & --- & ---\\	

\tiny{III}    	        &  \tiny{619.84 - $2.37 \times 10^{4}$}  & \tiny{$-2.82_{-0.20}^{+0.18}$}		& \tiny{---}  & \tiny{$1.02_{-0.06}^{+0.06}$} & \tiny{620 - 1140 } & \tiny{$-3.30_{-0.05}^{+0.12}$}& \tiny{584/(49, 3)} &\tiny{$1.18_{-0.21}^{+0.22}$}  \\	
\tiny{} & \tiny{$\rm t_b$ = $1570_{-67}^{+68}$}   	& \tiny{$1.43_{-0.05}^{+0.05}$} & \tiny{16425/(449, 5)}  	&\tiny{$1.05_{-0.09}^{+0.09}$} &  \tiny{4860 - $2.37 \times 10^{4}$}		& \tiny{$1.80_{-0.06}^{+0.05}$} &\tiny{1988/(155, 3)} &\tiny{$1.01_{-0.09}^{+0.09}$}   \\	

\tiny{IV}    	        & \tiny{$2.37 \times 10^{4}$ - $9.60 \times 10^{5}$}  & \tiny{$1.02_{-0.03}^{+0.03}$}& \tiny{---}	 &\tiny{$1.13_{-0.09}^{+0.09}$}	 & \tiny{$2.37 \times 10^{4}$ - $1.55 \times 10^{6}$} & \tiny{$0.90_{-0.10}^{+0.10}$} & ---& \tiny{$1.02_{-0.11}^{+0.12}$}   \\	
& \tiny{$\rm t_b$ = $3.95_{-1.28}^{+2.39}$ $\times10^{5}$}  & \tiny{$1.33_{-0.29}^{+0.13}$}	& \tiny{5053/(190, 5)} &\tiny{$1.32_{-0.24}^{+0.25}$}  & \tiny{$\rm t_b$ = $1.67_{-0.3}^{+0.4} \times 10^{5}$} & \tiny{$1.97_{-0.11}^{+0.12}$}  & \tiny{3301/(200, 5)}   & \tiny{$1.35_{-0.13}^{+0.13}$}\\ 
\tiny{IV $\rm-V^b$}   & \tiny{$9.60 \times 10^{5}$ - $7.90 \times 10^{6}$} & \tiny{$1.1_{-0.1}^{+0.1}$}  & \tiny{4031/(154, 3)} & \tiny{$1.05_{-0.26}^{+0.28}$ }& \tiny{$2.37 \times 10^{4}$ - $6.87 \times 10^{6}$} & \tiny{$1.22_{-0.03}^{+0.03}$} & \tiny{2468/(155, 3)} & \tiny{$1.21_{-0.1}^{+0.1}$}
\\
\hline
\end{tabular}
\label{tab:table_SED1}
\end{center}

We use power law 
or a smooth broken power-law (SBPL) \footnote{\begin{equation*}
Flux \propto \left[\left(\frac{t}{t_{\rm
b}}\right)^{\rm \alpha_1s} +\left(\frac{t}{t_{\rm
b}}\right)^{\rm \alpha_2s}\right]^{\rm -1/s} 
\end{equation*}
$\rm s$: smoothness parameter, $\rm \alpha_1$: index before break $\rm t_b$ and $\rm \alpha_2$: index after break  \citep{wang18:jet}, $\rm log(L)$: log-likelihood, n: number of data points and k: number of the free parameters in the fit. \\ $\rm b$: excluding $3 \times 10^5$ s - $2 \times 10^{6}$ s.
} function to model the XRT light curves for different segments using {\it python} module \sw{emcee}\citep{emcee}. All quoted errors on spectral parameters correspond to 16th and 84th percentiles.

\end{table*}

%% file: table_two_episodes.tex
\clearpage
\begin{table*}%
\caption{Bursts with two-episode emission and redshift detection}
\setlength{\tabcolsep}{2pt}

\begin{scriptsize}
\begin{center}
\begin{tabular}{ccccccccccc}
\\
\hline
\\
Trigger ID& z & $\rm T_{Ep.,1}$,   &    & $\rm F_{Ep.,1}$  & $\rm F_{peak,1}$ & $\rm E_{p,1}$ & $\rm \alpha_{1}$  & $\rm E_{iso, 1}$ & $\rm L_{iso,1}$ & Model$\rm _{Ep.,1}$  \\ 

&  &  $\rm T_{Ep.,2}$ & $\rm T_{q}$ & $\rm F_{Ep.,2}$ & $\rm F_{peak,2}$ & $\rm E_{p,2}$ & $\rm \alpha_{2}$ & $\rm E_{iso,2}$ & $\rm L_{iso,2}$ & Model$\rm _{Ep.,2}$ \\

&  &  ($\rm s$)  & ($\rm s$) & &    & ($\rm keV$) &  &  & &  \\

\hline
bn090328401 &$0.736$&$36.68$&--&$1.949_{-0.138}^{+0.136}$& $4.290_{-0.287}^{+0.862}$ &$655_{-65}^{+78}    $&$-0.96_{-0.03}^{+0.03}  $&$ 10.62   $&$  1.054 $& Band  \\ & &$ 18.36 $&$ 13.31 $&$ 0.117_{-0.014}^{+0.018} $& $0.410_{-0.104}^{+0.092}$  &$  96_{-17}^{+32}   $&$+1.26_{-0.21}^{+0.19}  $&$   0.37 $&$ 0.101  $& COMP   \\
\hline
bn091208410 &$1.063$&$ 4.33$&--&$0.392_{-0.041}^{+0.051}$& $0.751_{-0.105}^{+0.153}$ &$102_{-17}^{+28}    $&$+1.19_{-0.19}^{+0.17}  $&$  4.27   $&$  0.461 $& COMP  \\ & &$  5.85 $&$  3.58 $&$ 0.656_{-0.045}^{+0.052} $& $2.381_{-0.167}^{+0.193}$  &$ 129_{-15}^{+20}   $&$+1.15_{-0.10}^{+0.10}  $&$   1.14 $&$ 1.461  $& COMP   \\
\hline
bn100615083 &$1.398$&$25.51$&--&$0.369_{-0.066}^{+0.084}$& $0.869_{-0.184}^{+0.209}$ &$ 86_{-31}^{+18}    $&$-1.10_{-0.13}^{+0.29}  $&$  0.89   $&$  1.058 $& Band  \\ & &$ 18.37 $&$  3.07 $&$ 0.130_{-0.013}^{+0.016} $& $0.264_{-0.047}^{+0.062}$  &$  72_{-11}^{+17}   $&$-1.35_{-0.18}^{+0.16}  $&$   1.21 $&$ 0.321  $& COMP   \\ 
\hline
bn111228657 &$0.714$&$37.76$&--&$0.259_{-0.024}^{+0.027}$& $1.705_{-0.197}^{+0.268}$ &$ 48_{-4}^{+4}      $&$-1.63_{-0.07}^{+0.07}  $&$  1.01   $&$  0.389 $& Band  \\ & &$ 20.39 $&$ 19.46 $&$ 0.284_{-0.017}^{+0.017} $& $0.830_{-0.051}^{+0.095}$  &$  15_{-1}^{+1}     $&$+0.41_{-1.35}^{+3.70}$&$   0.84 $&$ 0.189  $& Band  \\
\hline
bn120711115 &$1.405$&$10.20$&--&$0.354_{-0.104}^{+0.163}$& $1.079_{-0.381}^{+0.982}$ &$389_{-131}^{+278}  $&$+0.41_{-0.62}^{+0.42}  $&$  6.85   $&$  1.329 $& COMP  \\ & &$ 65.16 $&$ 50.18 $&$ 5.160_{-0.133}^{+0.105} $& $15.075_{-1.607}^{+1.842}$ &$1123_{-88}^{+58}   $&$-0.95_{-0.01}^{+0.02}  $&$ 172.27 $&$18.576  $& Band   \\
\hline
bn120716712 &$2.486$&$ 7.14$&--&$0.236_{-0.036}^{+0.051}$& $0.957_{-0.123}^{+0.125}$ &$152_{-37}^{+74}    $&$+1.10_{-0.26}^{+0.21}  $&$  3.49   $&$  4.841 $& COMP  \\ & &$ 47.95 $&$169.47 $&$ 0.263_{-0.024}^{+0.034} $& $0.568_{-0.046}^{+0.085}$  &$ 113_{-11}^{+10}   $&$-0.87_{-0.09}^{+0.11}  $&$  18.30 $&$ 2.873  $& Band   \\
\hline
bn131108024 &$2.400$&$12.24$&--&$0.236_{-0.024}^{+0.031}$& $0.868_{-0.129}^{+0.161}$ &$132_{-22}^{+34}    $&$+1.03_{-0.19}^{+0.17}  $&$  2.30   $&$  4.028 $& COMP  \\ & &$ 54.09 $&$219.13 $&$ 0.176_{-0.035}^{+0.084} $& $0.403_{-0.112}^{+0.250}$  &$ 410_{-139}^{+133} $&$-0.46_{-0.23}^{+0.35}  $&$  24.44 $&$ 1.870  $& Band   \\
\hline
bn140304849 &$5.283$&$36.74$&--&$0.146_{-0.038}^{+0.034}$& $0.404_{-0.100}^{+0.160}$ &$548_{-307}^{+1043} $&$+1.41_{-0.19}^{+0.13}  $&$  8.78   $&$ 12.476 $& COMP  \\ & &$ 70.40 $&$148.48 $&$ 0.084_{-0.008}^{+0.009} $& $0.333_{-0.069}^{+0.091}$  &$  94_{-13}^{+18}   $&$+1.03_{-0.16}^{+0.18}  $&$  29.07 $&$10.284  $& COMP   \\
\hline
bn140512814 &$0.725$&$13.25$&--&$0.756_{-0.103}^{+0.121}$& $2.297_{-0.648}^{+0.710}$ &$627_{-192}^{+344}  $&$+1.09_{-0.07}^{+0.06}  $&$  3.81   $&$  0.544 $& COMP  \\ & &$ 67.27 $&$ 83.97 $&$ 0.555_{-0.045}^{+0.054} $& $2.941_{-0.516}^{+0.778}$  &$ 436_{-59}^{+78}   $&$+1.16_{-0.04}^{+0.04}  $&$   5.12 $&$ 0.696  $& COMP   \\
\hline
bn151027166 &$0.810$&$25.50$&--&$0.209_{-0.016}^{+0.021}$& $1.125_{-0.098}^{+0.097}$ &$135_{-20}^{+29}    $&$+1.31_{-0.11}^{+0.11}  $&$  1.32   $&$  0.351 $& COMP  \\ & &$ 31.62 $&$ 71.68 $&$ 0.497_{-0.078}^{+0.092} $& $1.270_{-0.349}^{+0.631}$  &$ 538_{-173}^{+271} $&$+1.42_{-0.07}^{+0.05}  $&$   2.71 $&$ 0.396  $& COMP   \\
\hline
bn180728728 &$0.117$&$ 4.08$&--&$0.313_{-0.041}^{+0.044}$& $0.536_{-0.091}^{+0.081}$ &$ 14_{-4}^{+2}      $&$+0.21_{-1.98}^{+\infty}$&$  0.02   $&$  0.002 $& Band  \\ & &$  9.02 $&$  7.17 $&$ 6.392_{-0.010}^{+0.012} $& $19.794_{-0.501}^{+0.522}$ &$  81_{-1}^{+1}     $&$-1.53_{-0.01}^{+0.01}  $&$   0.18 $&$ 0.068  $& Band   \\
\hline
\end{tabular}
 \end{center}
 $\rm F_{Ep.,1}$, $\rm F_{Ep.,2}$, $\rm F_{Ep.,1}$ and $\rm F_{Ep.,2}$ : in units of $\rm 10^{-6}$ $\rm erg~cm^{-2}~s^{-1}$;
 $\rm E_{iso, 1}$ and   $\rm E_{iso,2}$  : in units of $\rm 10^{52}$ $\rm erg$; $\rm L_{iso,1}$ and $\rm L_{iso,2}$ in units of $\rm 10^{52}$ $\rm erg$  $\rm s^{-1}$
\end{scriptsize}
\label{tab:two_episode_grbfits}
\end{table*}

%% file: table_UVOT.tex

\begin{table*}
\begin{scriptsize}
\begin{center}
\setlength{\tabcolsep}{2pt}

\caption{Log of UVOT observations $\&$ photometry of the GRB 190829A afterglow. No correction for galactic extinction is applied.}
\label{table:UVOT}

\begin{tabular}{@{}llllll@{}}
\toprule
Filter & $\rm T_{i}$ ($\rm s$) & $\rm T_{s}$ ($\rm s$) & Magnitude& Flux density ($\rm mJy$) \\ \hline
white & 106.3 & 256.1 & 19.51 $\pm$ 0.14 & 0.036 $\pm$ 0.005\\
white & 545.8 & 565.6 & 19.12 $\pm$ 0.29 & 0.051 $\pm$ 0.014\\
white & 721.6 & 741.3 & 18.87 $\pm$ 0.24 & 0.064 $\pm$ 0.014 \\
white & 870.6 & 1020.0 & 17.57 $\pm$ 0.04& 0.213 $\pm$ 0.008  \\
white & 1173.5 & 1366.5 & 16.88 $\pm$ 0.06& 0.401 $\pm$ 0.022 \\
white & 1519.3 & 1701.6 & 16.96 $\pm$ 0.08 & 0.373 $\pm$ 0.027 \\
white & 6079.3 & 6279.0 & 18.21 $\pm$ 0.05& 0.118 $\pm$ 0.005  \\
white & 10724.2 & 11631.1 & 18.85 $\pm$ 0.04& 0.065 $\pm$ 0.002 \\
white & 172223.6 & 173130.6 & 19.53 $\pm$ 0.06 & 0.035 $\pm$ 0.002 \\ \hline
b & 521.3 & 541.1 &  $>$ 18.76 & $>$ 0.127\\
b & 696.3 & 716.1 &  $>$ 18.76 & $>$ 0.127 \\
b & 1148.7 & 1342.2 & 17.24 $\pm$ 0.11& 0.516 $\pm$ 0.052  \\
b & 1495.0 & 1687.4 & 17.52 $\pm$ 0.16& 0.399 $\pm$ 0.059  \\
b & 5874.5 & 6074.3 & 18.43 $\pm$ 0.09& 0.172 $\pm$ 0.014 \\
b & 7310.3 & 7438.8 & 18.84 $\pm$ 0.30& 0.118 $\pm$ 0.032  \\
b & 99230.7 & 171611.5 & 19.80 $\pm$ 0.22 & 0.049 $\pm$ 0.010  \\
b & 171615.3 & 172218.5 & 19.54 $\pm$ 0.12& 0.062 $\pm$ 0.007 \\ \hline
u & 264.3 & 514.1 & 19.43 $\pm$ 0.23 & 0.025 $\pm$ 0.005 \\
u & 671.6 & 864.6 & 18.64 $\pm$ 0.34 & 0.052 $\pm$ 0.016  \\
u & 1123.8 & 1317.6  & 17.27 $\pm$ 0.16& 0.183 $\pm$ 0.027 \\
u & 1469.8 & 1662.5 & 17.77 $\pm$ 0.29& 0.115 $\pm$ 0.031  \\
u & 5669.3 & 5869.1 & 18.25 $\pm$ 0.11& 0.074 $\pm$ 0.007  \\
u & 7105.2 & 7304.9 & 18.56 $\pm$ 0.24& 0.056 $\pm$ 0.012  \\
u & 24564.2 & 24650.4 &  $>$ 18.40 & $>$  0.065\\
u & 98317.4 & 109397.7 & 19.35 $\pm$ 0.18& 0.027 $\pm$ 0.004  \\ 
u & 264490.8 & 271341.2 & 19.89 $\pm$ 0.22& 0.016 $\pm$ 0.003  \\\hline
v & 597.5 & 617.3 &  $>$ 17.72 &  $>$ 0.297 \\
v & 771.3 & 791.0 & 17.22 $\pm$ 0.27 & 0.471 $\pm$ 0.117  \\
v & 1050.3 & 1242.8 & 16.02 $\pm$ 0.10 & 1.421 $\pm$ 0.131   \\
v & 1396.2 & 1589.4 & 15.81 $\pm$ 0.10 & 1.724 $\pm$ 0.159   \\
v & 5054.5 & 5254.2 & 17.01 $\pm$ 0.08 & 0.571 $\pm$  0.042 \\
v & 6490.7 & 6690.4 & 17.36 $\pm$ 0.10 & 0.414 $\pm$ 0.038   \\
v & 17608.5 & 18515.3 & 18.49 $\pm$ 0.11 & 0.146 $\pm$ 0.015   \\
v & 45033.9 & 45941.0 & 18.59 $\pm$ 0.11 & 0.133 $\pm$ 0.013  \\
v & 167595.5 & 179056.0 & 18.82 $\pm$ 0.12 & 0.108 $\pm$ 0.012   \\ \hline
uvw1 & 647.4 & 1119.1 & 18.48 $\pm$ 0.36 & 0.040 $\pm$ 0.013   \\ 
uvw1 & 1273.5 & 1638.2 & $>$ 18.15 & $>$ 0.054\\
uvw1 & 5464.5 & 5664.2 & 19.26 $\pm$ 0.32 & 0.019 $\pm$ 0.006   \\
uvw1 & 6900.5 & 7100.3 & 18.99 $\pm$ 0.32 & 0.025 $\pm$ 0.007   \\
uvw1 & 23658.1 & 24557.8 & 19.74 $\pm$ 0.26 & 0.012 $\pm$ 0.003   \\
uvw1 & 87378.3 & 104975.0 & $>$ 20.48 & $>$ 0.006\\
uvw1 & 259190.3 & 270886.2 & 20.38 $\pm$ 0.33 & 0.007 $\pm$ 0.002   \\\hline
uvw2 & 4849.7 & 5049.5 & 19.19 $\pm$ 0.28 & 0.016 $\pm$ 0.004   \\
uvw2 & 6285.7 & 6485.5 & 19.11 $\pm$ 0.28 & 0.017 $\pm$  0.004 \\
uvw2 & 11637.3 & 12112.8 & 20.01 $\pm$ 0.33 & 0.007 $\pm$ 0.002   \\
uvw2 & 27922.6 & 30389.2 & 19.71 $\pm$ 0.29 & 0.010 $\pm$ 0.003   \\
uvw2 & 33673.8 & 40012.9 & 20.14 $\pm$ 0.26 & 0.007 $\pm$ 0.002  \\
uvw2 & 166688.4 & 173821.9 & $>$ 20.62 & $>$ 0.004 \\ \hline
uvm2 & 5259.3 & 5459.1 & $>$ 19.33 &  $>$ 0.014 \\
uvm2 & 6695.5 & 6895.2 & $>$ 19.22 &  $>$ 0.016  \\
uvm2 & 18520.3 & 18914.0 & $>$ 19.63 &  $>$ 0.011  \\
uvm2 & 22751.2 & 23651.0 & 20.39 $\pm$ 0.36 & 0.005 $\pm$ 0.002   \\
uvm2 & 86471.6 & 103671.5 & 20.00 $\pm$ 0.19 & 0.008 $\pm$ 0.001   \\
uvm2 & 179061.3 & 259183.4 & 20.17 $\pm$ 0.29 & 0.007 $\pm$  0.002 \\
uvm2 & 269079.1 & 269978.9 & 20.38 $\pm$ 0.34 & 0.005 $\pm$ 0.002 \\\hline
\end{tabular}
 \end{center}

\end{scriptsize}
\end{table*}